\documentstyle[aps]{revtex}

\newcommand{\beq}{\begin{equation}}
\newcommand{\eeq}{\end{equation}}
\newcommand{\bear}{\begin{eqnarray}}
\newcommand{\ear}{\end{eqnarray}}
\newcommand{\earn}{\nonumber \end{eqnarray}}

\newcommand{\nn}{\nonumber \\}

\newcommand{\mds}{m_{\rm DS}}
\newcommand{\gsim}{\mathop{\lefteqn{\raise.9pt\hbox{$>$}}
\raise-3.7pt\hbox{$\sim$}}}
\newcommand{\lsim}{\mathop{\lefteqn{\raise.9pt\hbox{$<$}}
\raise-3.7pt\hbox{$\sim$}}}
\arraycolsep=\mathsurround

\begin{document}

\title{Stress-energy of a quantized scalar field in static
wormhole spacetimes}

\author{Arkadii A. Popov\thanks{Email address: popov@kspu.kcn.ru}}

\address{Department of Geometry, Kazan State Pedagogical University,
Mezhlauk 1 Steet, Kazan 420021, Russia}

\maketitle

\begin{abstract}
An analytical approximation of $\langle T_{\mu\nu} \rangle$ for a quantized 
scalar field in a static spherically symmetric  spacetime with a topology 
$S^2 \times R^2$ is obtained. The gravitational background is assumed 
slowly varying. The scalar field is assumed to be both massive and massless, 
with an arbitrary coupling $\xi$ to the scalar curvature and in a zero 
temperature vacuum state. It is demonstrated that for some values 
of curvature coupling the stress-energy has the properties needed 
to support the wormhole geometry.

\vskip12pt\noindent
{\normalsize PACS number(s): 04.62.+v, 04.70.Dy}
\end{abstract}


\section{Introduction}

In recent years Lorentzian wormholes have been of some interest as 
nontrivial geometric and  physical objects.  The topology of such spacetimes 
is $R^2 \times S^2$. The characteristic feature of wormhole geometry
is the existence of a sphere of minimal square which cannot be 
contracted to the point. The possibility of the  existence of
a static spherically symmetrical traversable wormhole as 
a topology-nontrivial solution of the Einstein equations was 
first studied by Morris and Thorne \cite{MT}. They found 
that the matter which threads the wormhole throat should have 
unusual properties. In particular, the radial tension of the matter 
should exceed its density both locally at the throat \cite{MT}
and integrally along the radial direction \cite{MTY}. As an example
of such matter they suggested the Casimir vacuum between
conducting spherical plates. 
Violations of the energy conditions in static wormholes 
have been analyzed in detail by a number of authors (see, for example,  
\cite{hocvis}). Recently Barcel\'o  and Visser \cite{BW1,BW2} 
demonstrated that a scalar field with a positive 
curvature coupling violates all the standard energy conditions.
They  found  the entire branch of traversable wormhole solutions 
for gravity plus a nonminimally coupled massless scalar field. 
Solutions describing primordial wormholes are obtained 
for ${\cal N}=4$ SU(N) super Yang-Mills theory in \cite{Od2}.
The possibility of the existence of wormhole solutions has been investigated 
in Brans-Dicke theory \cite{Agnese,Nandi,Anch}, higher 
derivative gravity \cite{Hoch}, and Gauss-Bonnet theory \cite{Bhawal}.

Within the framework of general relativity as an example 
of the matter that supports the wormhole 
geometry one can consider the vacuum of quantized fields. 
This approach gives the possibility of describing
the wormhole metric as a self-consistent solution of the semiclassical
Einstein field equations \cite{Sush,HPS}. The principal problem encountered 
in this method is the impossibility of calculating the functional 
dependence of the stress-energy tensor of quantized fields on 
the metric tensor. Nevertheless, in some cases approximate 
expressions can be obtained 
\cite{Page,BO,BOP,FZ,FZ1,FZ2,FZ3,FZ4,FSZ,Avra1,Avra2,Avra3,Khat1,Khat2,Mat,AHS}.
In above mentioned work \cite{Sush,HPS} the approximations
of Frolov-Zel'nikov \cite{FZ4} and Anderson-Hiscock-Samuel \cite{AHS}
were used. But as noted in \cite{PS} in spacetime that is 
a direct product of the Minkowski plane and a two-dimensional
sphere of fixed radius these approximations are not applicable. 
The approximate expressions for the electromagnetic and
neutrino stress-energy tensor obtained 
by Khatsymovsky \cite{Khat1,Khat2} do not have this deficiency.
In this work the results were obtained by the WKB method, and,
as the zero order of the expansions was used, the exact solutions 
of the mode equations were obtained in spacetime with the metric 
$ds^2=-f_0dt^2+dl^2+r_0^2(d\theta^2+\sin^2\theta\, d\varphi^2)$
 ($f_0$  and $r_0$ are constants),
i.e., in the above mentioned spacetime.
It is well known that the vacuum expectation value of the stress-energy
tensor of quantized fields depends on the topology of spacetime. 
In this connection the approach of Khatsymovsky seems more 
natural in wormhole spacetimes.

In this paper an approximate expression for the stress-energy tensor 
of a quantized scalar field in static spherically 
symmetric  spacetimes with topology $R^2 \times S^2$ is calculated. 
The Anderson-Hiscock-Samuel approach \cite{AHS} is used
but the zeroth order of the WKB solution of the mode equation
is redefined by analogy with \cite{Khat1,Khat2} and mode sums
are computed exactly as in \cite{Sushkov2,PS}. 

The units $\hbar=c=1$ are used throughout the paper. 


\section{An unrenormalized expression for  
$\left< T^{\mu}_{\nu} \right>$}

In this section the Euclidean space approach is used to calculate
an unrenormalized expression for $\langle T^{\mu}_{\nu} \rangle$ for 
a scalar field in a static spherically symmetric spacetime with 
topology $S^2 \times R^2$. 
The metric for a general static spherically symmetric spacetime 
when continued analytically into Euclidean space can be written as
  \beq\label{metric}
  ds^2= f(\rho)d\tau^2+d\rho^2+r^2(\rho)(d\theta^2+\sin^2\theta\, d\varphi^2).
  \eeq
We assume that $ - \infty < \rho < \infty$, $f$ and $r$ are
arbitrary functions of $\rho$, and $\tau$ is the Euclidean time ($\tau = it,$ 
where $t$ is the coordinate corresponding to the timelike Killing vector
which always exists in a static spacetime).

$\langle T^{\mu}_{\nu} \rangle$ of a quantized scalar field $\phi$ 
can be computed using the method of point
splitting from the Euclidean Green's function $G_E(x,\tilde x)$  as follows
\cite{AHS}
 \bear \label{Tmnunren}
        \langle T^{\mu}_{\nu} \rangle_{unren}&=&
        \left( {1/2-\xi} \right) (g^{\mu \tilde \alpha} 
        G_{E; \tilde \alpha \nu} + 
        g_{\nu}^{\tilde \alpha} G_{E;}{}^{\mu}{}_{\tilde \alpha}) +
        (2 \xi-1/2) \delta^{\mu}_{\nu} g^{\sigma \tilde \alpha}
        G_{E; \sigma \tilde \alpha}
        -\xi (G_{E;}{}^{\mu}{}_{\nu} \nonumber \\ &&
        +g^{\mu \tilde \alpha}g^{\tilde \beta}_{\nu}
        G_{E;\tilde \alpha \tilde \beta})
        +2 \xi \delta^{\mu}_{\nu} (m^2 + \xi R) G_E
        +\xi (R^{\mu}_{\nu}-\delta^{\mu}_{\nu} R / 2)  
        G_E-\delta^{\mu}_{\nu} m^2 G_E/2,
        \ear
where $m$ is the mass of the scalar field, $\xi$ is its coupling to the scalar
curvature $R$, and $g^{\tilde \alpha}_{\beta}$ is the bivector of parallel 
transport of a vector at $\tilde x$ to one at $x$.

In \cite{AHS} the form of $G_E(x,\tilde x)$ was derived for a scalar field
in a static spherically symmetric spacetime when the field is in the zero 
temperature vacuum state defined with respect to the timelike Killing vector
\bear \label{GE}
      G_E(x;\tilde x)=\frac{1}{4 \pi^2}\int \nolimits_{0}^{\infty } d \omega
      \cos [\omega (\tau - \tilde \tau)] \sum \limits_{l=0}^{\infty}
      (2l+1) P_l (\cos \gamma ) \ C_{\omega l} \ p_{\omega l}(\rho_<) 
      \ q_{\omega l}(\rho_>),
        \ear
where $P_l$ is a Legendre polynomial, $\cos \gamma \equiv \cos \theta \cos 
\tilde \theta+\sin \theta \sin \tilde \theta \cos (\varphi -\tilde \varphi)$,
$C_{\omega l}$ is a normalization constant,
$\rho_<$ and $\rho_>$ represent the lesser and greater of $\rho$ and 
$\tilde \rho$, respectively, and the modes $p_{\omega l}(\rho)$ and 
$q_{\omega l}(\rho)$ obey the equation
   \bear\label{modeeqn} \left\{ { 
   \frac{d^2}{d\rho^2}+\left[\frac{1}{2f}\frac{df}{d\rho}
   +\frac{1}{r^2}\frac{dr^2}{d\rho}\right]\frac{d}{d\rho}
   -\left[\frac{\omega^2}{f}+\frac{l(l+1)}{r^2}+m^2+\xi R\right]}\right\}
   \left\{ {\begin{array}{l}p_{\omega l}\\ q_{\omega l}\end{array}}\right\}=0.
   \ear
They also satisfy the Wronskian condition
        \beq\label{wronskian}
        C_{\omega l}\left[p_{\omega l}\frac{dq_{\omega l}}{d\rho}-
        q_{\omega l}\frac{dp_{\omega l}}{d\rho}\right]=\frac{-1}{r^2f^{1/2}}.
        \eeq
After the substitution
        \beq\label{modes}
        \begin{array}{l}
        \displaystyle
        p_{\omega l}=\frac1{\sqrt {2 r^2 W}}
        \exp \left\{ \int^\rho W f^{-1/2} d\rho \right\}, \\
        \\
        \displaystyle
        q_{\omega l}=\frac1{\sqrt{2 r^2 W}}
        \exp\left\{- \int^\rho W f^{-1/2} d\rho \right\},
        \end{array}
        \eeq
it is easy to see that the Wronskian condition (\ref{wronskian}) is obeyed if 
       \beq
        C_{\omega l}=1 
         \eeq 
and the mode equation (\ref{modeeqn})
gives the following equation for $W(\rho)$:
     \beq \label{W2}
     W^2=\omega^2+\frac{f}{r^2}\left[
     {l(l+1)+2\xi+m^2r^2} \right]
     + { \frac{f'}{8}\frac{(W^2)'}{W^2}
     +\frac{f}{4}\frac{(W^2)''}{W^2}
     -\frac{5f}{16}\frac{(W^2)'^2}{W^4} }+V,
     \eeq
where
     \bear
     V&=&f\left( { \frac{(r^2)''}{2r^2}+\frac{f'(r^2)'}{4fr^2}
     -\frac{(r^2)'^2}{4r^4}} \right) 
    + \xi f \left( {-\frac{f''}{f}-2\frac{(r^2)''}{r^2}
    +\frac{f'^2}{2f^2} +\frac{(r^2)'^2}{2r^4}-\frac{f'(r^2)'}{fr^2}
     } \right).
     \ear    
The prime denotes the derivative with respect to $\rho$.

Equation (\ref{W2}) can be solved iteratively when 
the metric functions $f(\rho)$ and $r^2(\rho)$ are varying slowly,
that is,
     \beq \label{lwkb}
     \lambda_{WKB}=L_{\star} /L \ll 1,
     \eeq
where 
      \beq
      L_{\star} =\left[ m^2+ \frac{2\xi}{r^2} \right]^{-1/2},
      \eeq
and $L$ is a characteristic scale of variation of the metric
functions:
       \beq \label{Lm}
       L^{-1}= \max \left \{ \left| \left[ \ln(fr^2) \right]'  \right|, \
       \left| \left[ \ln(fr^2) \right]''  \right|^{1/2}, \
       \left| \left[ \ln(fr^2) \right]'''  \right|^{1/3}, \
       \dots  \right \} .
       \eeq
The zeroth-order WKB solution of Eq. (\ref{W2}) corresponds 
to neglecting terms with derivatives in this equation
       \beq
       W^2=U_0,
       \eeq
where
        \bear
        U_0&=&\omega^2+\frac{f}{r^2}\left(l+\frac12\right)^2
        +\frac{f}{r^2}\mu^2, \nn
       \mu^2&=&m^2r^2+2\xi-\frac14.
        \ear
Let us stress that $U_0$ is the exact solution of Eq. (\ref{W2}) in
a spacetime with metric 
$ds^2=f_0d\tau^2+d\rho^2+r_0^2(d\theta^2+\sin^2\theta\, d\varphi^2)$,
where $f_0$ and $r_0$ are constants.

Below as in Ref. \cite{PS}  it is assumed that
       \bear
       \mu^2>0
       \ear
or
       \bear
       r^2>\frac{1-8\xi}{4m^2}.
       \ear
The second-order solution is
        \beq\label{wkbsol}
        W^2=U_{0}+V+\frac{f'}{8} \frac{U_{0}{}'}{U_{0}}
        +\frac{f}{4} \frac{U_{0}{}''}{U_{0}}
        -\frac{5f}{16} \frac{U_{0}{}'^2}{U_{0}^2}.
        \eeq
Now we can rewrite expression (\ref{Tmnunren}) using expressions 
(\ref{GE}) and (\ref{modes}) and then suppose 
$\rho=\tilde \rho, \theta=\tilde \theta, \phi=\tilde \phi$. The superficial
divergences in the sums over $l$ that appear in this case can be 
removed as in Refs.\cite{CH,AHS}:
     \bear \label{Ttt}
     \left< T^t_t \right>_{unren}&=&\left[\frac12 g^{t \tilde t}-\frac{\xi}{f}
     -\xi f (g^{t \tilde t})^2-\xi(g^{\rho \tilde t})^2 \right]\frac{f}{r^2}
     \frac{\partial^2 }{\partial \varepsilon^2} B_1
     +\left( 2\xi-\frac12 \right)g^{\rho \tilde \rho} B_2 
     +\frac 1{r^2}\left[ \xi f (g^{t \tilde \rho})^2 \right. \nn &&
      \left.+\left(2\xi-\frac12  \right)  \right]B_3
     +\xi\left[ -\frac{f'}{2f}-\frac{ff'}{2}(g^{t \tilde t})^2
     +f\left( -\frac{f'}{2f}-\frac{(r^2)'}{r^2} \right)(g^{t \tilde \rho})^2 
     \right]B_4 \nn
     &&+\left[ \xi f\left(- \frac{1}{4r^2}+m^2+\xi R \right)
    (g^{t\tilde \rho})^2
    -\left( 2\xi-\frac12 \right)\frac{1}{4r^2}+\left( 2\xi-\frac12 \right) 
    (m^2+\xi R)  \right. \nn && \left.
     +\xi R^t_t  \right] B_1 
     +i \xi f'g^{t\tilde t}g^{t\tilde\rho}\sqrt{\frac{f}{r^2}}
      \frac{\partial}{\partial\varepsilon}B_1+i \xi \left[2 g^{t\tilde \rho}
     +2 f g^{t\tilde t} g^{\rho\tilde t} \right] \sqrt{\frac{f}{r^2}}
     \frac{\partial}{\partial\varepsilon}B_4, 
     \ear
     \bear \label{Trr}
     \left< T^{\rho}_{\rho} \right>_{unren}&=&\left[\left(2\xi-\frac12\right)
      g^{t \tilde t}+\frac{\xi}{f}+ \xi (g^{\rho \tilde t})^2
     +\frac{\xi}{f}(g^{\rho \tilde \rho})^2 \right] \frac{f}{r^2}
     \frac{\partial^2 }{\partial \varepsilon^2} B_1
     +\frac12 g^{\rho \tilde \rho} B_2 \nn
     &&+\frac 1{r^2}\left[\xi- \xi (g^{\rho \tilde \rho})^2-\frac12  
     \right]B_3 +\left[\xi\left( \frac{f'}{2f}+\frac{(r^2)'}{r^2}\right)
     (1+(g^{\rho \tilde \rho})^2)
     +\frac{\xi f'}{2}(g^{\rho \tilde t})^2 \right]B_4 \nn
     &&+\left[ \xi (1+(g^{\rho \tilde \rho})^2) \left( \frac{1}{4r^2}-m^2
    -\xi R \right)
     -\left( 2\xi-\frac12 \right)\frac{1}{4r^2}
    +\left( 2\xi-\frac12 \right)(m^2+\xi R)
     \right. \nn && \left.+\xi R^{\rho}_{\rho}  \right] B_1 
     +i \xi \frac{f'}{f}g^{\rho\tilde \rho}g^{t\tilde \rho}
     \sqrt{\frac{f}{r^2}} \frac{\partial}{\partial\varepsilon}B_1
    +i \xi \left[ 2 g^{t\tilde \rho}+2 g^{\rho\tilde \rho} g^{\rho\tilde t} 
    \right] 
    \sqrt{\frac{f}{r^2}}\frac{\partial}{\partial\varepsilon}B_4, 
    \ear
    \bear \label{Tthth}
    \left< T^{\theta}_{\theta} \right>_{unren}&=&\left(2\xi-\frac12\right)
    g^{t \tilde t} \frac{f}{r^2} \frac{\partial^2 }{\partial \varepsilon^2} B_1
    +\left(2\xi-\frac12 \right) g^{\rho \tilde \rho} B_2 +\frac {2\xi}{r^2}B_3 
    -\frac{\xi}{2}\frac{(r^2)'}{r^2}B_4 \nn
    &&+\left[-\frac{2\xi}{r^2}
   +\left( 2\xi-\frac12 \right)(m^2+\xi R)+\xi R^{\theta}_{\theta}  \right] B_1
   +i 2\left(2\xi-\frac12 \right) g^{t\tilde \rho}
   \sqrt{\frac{f}{r^2}}\frac{\partial}{\partial\varepsilon}B_4, 
   \ear
where
     \bear\label{B1}
     B_1&=&\frac{1}{4\pi^2}\int \nolimits_{0}^{\infty}du \cos(u \varepsilon)
     \sum \limits_{l=0}^{\infty}\frac1{r^2}
     \left[ \sqrt{\frac{f}{r^2}} \frac{(l+1/2)}{W}-1 \right],
      \ear
     \bear
     B_2&=&\frac{1}{4\pi^2}\int \nolimits_{0}^{\infty}du \cos(u \varepsilon)
     \sum \limits_{l=0}^{\infty} \frac1{r^4}
     \left[-\left(l+ \frac{1}{2} \right) \sqrt{\frac{r^2}{f}}W
    +\left(l+ \frac{1}{2} \right)\frac{(r^2)'{}^2}{4r^2}\sqrt{\frac{f}{r^2}} 
    \frac{1}{W} 
    \right.\nn
     &&+\left(l+ \frac{1}{2} \right)\frac{(r^2)'}{4}\sqrt{\frac{f}{r^2}}
    \frac{(W^2)'}{W^3}
    +\left(l+ \frac{1}{2} \right)\frac{r^2}{16}\sqrt{\frac{f}{r^2}}
    \frac{(W^2)'{}^2}{W^5}
     +\left(l+ \frac{1}{2} \right)^2
     +\frac{(u^2+\mu^2)}{2}  \nn
     &&+\frac{r^2}{2f}V
     -\frac{(r^2)'{}^2}{4r^2}
    +\frac{r^2 f'}{16f}\left( \frac{f}{r^2} \right)' \left( \frac{r^2}{f} 
    \right)
     -\frac{(r^2)'}{4}\left( \frac{f}{r^2} \right)' \left( \frac{r^2}{f} 
     \right)
    +\frac{r^2}{8}\left( \frac{f}{r^2} \right)'' \left( \frac{r^2}{f} \right)
    \nn &&\left.
     -\frac{7r^2}{32}\left( \frac{f}{r^2} \right)'^{{}2} 
     \left( \frac{r^2}{f} \right)^{-2} \right],
     \ear
     \bear
     B_3&=&\frac{1}{4\pi^2}\int \nolimits_{0}^{\infty}du \cos(u \varepsilon)
     \sum \limits_{l=0}^{\infty}\frac1{r^2}
     \left[ \sqrt{\frac{f}{r^2}}\frac{(l+1/2)^3}{W}-\left(l+\frac12 \right)^2
     +\frac{(u^2+\mu^2)}{2}+\frac{r^4}{8f}\left( \frac{f}{r^2}
    \right)''\right. \nn
     &&\left.+\frac{r^4f'}{16f^2}\left( \frac{f}{r^2} \right)'  
    -\frac{5r^4}{32f}\left( \frac{f}{r^2} \right)'^2+\frac{r^2}{2f}V  \right],
    \ear
    \bear\label{B4}
    B_4&=&\frac{1}{4\pi^2}\int \nolimits_{0}^{\infty}du \cos(u \varepsilon)
    \sum \limits_{l=0}^{\infty}\frac1{\left| r \right|^3 }
    \left[ \left( l+\frac12 \right)\frac{(r^2)'^2}{4r^2} \sqrt{\frac{f}{r^2}}
    \frac{1}{W}-\left( l+\frac12 \right)\frac{\left| r \right| }{4}
      \sqrt{\frac{f}{r^2}}\frac{(W^2)'}{W^3}\right.\nn
      &&\left.+\frac{(r^2)'}{4\left| r \right|}
      +\frac{\left| r \right| f'}{4f}\right], 
       \ear
      \beq
      \varepsilon=\sqrt{\frac{f}{r^2}}(\tau-\tilde \tau).
      \eeq

\section{Use of the WKB approximation in evaluating
$ \langle T^{\mu}_{\nu} \rangle $}

The approximate (of second WKB order) expressions for the quantities $B_1$, 
$B_2$, $B_3$, and $B_4$ are obtained by substituting the WKB expansion
of $W^2$ [Eq. (\ref{W2})] into Eqs. (\ref{B1})-(\ref{B4}):
     \bear\label{B1_}
     B_1&=&\
\frac{1}{4\pi^2}\left\{
{\frac {1}{r^2}} S^0_0(\varepsilon,\mu) 
-{\frac {V}{2f}} S^0_1(\varepsilon,\mu) 
-\frac{r^2}{16 f^2} \left[f'\left( \frac{f}{r^2} \right)' \right. \right.\nn
&&\left.+2 f\left( \frac{f}{r^2} \right)''\right] S^1_2(\varepsilon,\mu) 
-\frac{r^2}{16f^2}
\left[2f\left( \mu^2 \frac{f}{r^2} \right)''
+f'\left( \mu^2 \frac{f}{r^2} \right)'\right] S^0_2(\varepsilon,\mu) \nn &&
+{\frac {5r^4}{32f^2}}\left( \frac{f}{r^2} \right)'^{2} S^2_3(\varepsilon,\mu) 
+\frac{5 r^4}{16 f^3}
f\left( \frac{f}{r^2} \right)'\left( \mu^2 \frac{f}{r^2} \right)' 
 S^1_3(\varepsilon,\mu) \nn && \left. 
+\frac{5 r^4}{32 f^3} f\left( \mu^2 \frac{f}{r^2} \right)'^{2}
 S^0_3(\varepsilon,\mu) 
\right\} +O\left(\frac{r^2}{L^4}\right),
      \ear
     \bear\label{B2_}
B_2&=&\frac{1}{4\pi^2} \left\{
-\frac{1}{r^4} S^0_{-1}(\varepsilon,\mu) 
-{\frac {1}{4 f r^6}}\left[-\left( r^2 \right)'^{2}f+2V r^4\right] 
S^0_0(\varepsilon,\mu) \right. \nn && 
-\frac {1}{16 f^2 r^2}\left[2 f r^2\left(\frac{f}{r^2}\right)'' 
-4f\left( r^2 \right)'\left( \frac{f}{r^2} \right)' 
+r^2f'\left( \frac{f}{r^2} \right)'\right] S^1_1(\varepsilon,\mu) 
\nn && -\frac {1}{16 f^2 r^2 }
\left[r^2f'\left( \mu^2 \frac{f}{r^2} \right)' 
+2fr^2\left(\mu^2\frac{f}{r^2}\right)'' 
-4f\left( r^2 \right)'\left( \mu^2\frac{f}{r^2} \right)' 
\right] S^0_1(\varepsilon,\mu)  \nn && 
+\frac {7 r^2}{32 f^2}\left( \frac{f}{r^2} \right)'^{2} S^2_2(\varepsilon,\mu)
+\frac {7r^2}{16 f^2 }
\left( \frac{f}{r^2} \right)'\left( \mu^2\frac{f}{r^2} \right)'
 S^1_2(\varepsilon,\mu) \nn && \left.
+\frac {7r^2}{32f^2 }
\left( \mu^2\frac{f}{r^2} \right)'^{2}
S^0_2(\varepsilon,\mu) \right\} 
 +O\left(\frac{1}{L^4}\right),
     \ear
     \bear\label{B3_}
     B_3&=&
\frac{1}{4\pi^2} \left\{
{\frac {1}{r^2}} S^1_0(\varepsilon,\mu) 
-{\frac {V}{2f}} S^1_1(\varepsilon,\mu) 
-\frac{r^2}{16 f^2}\left[f'\left( \frac{f}{r^2} \right)' \right.\right. 
\nn && \left.
+2 f\left( \frac{f}{r^2} \right)''\right] S^2_2(\varepsilon,\mu)
-\frac{r^2}{16f^2}
\left[2f\left( \mu^2 \frac{f}{r^2} \right)''
+f'\left( \mu^2 \frac{f}{r^2} \right)'\right] S^1_2(\varepsilon,\mu) 
\nn &&+{\frac {5r^4}{32f^2}}\left( \frac{f}{r^2} \right)'^{2} 
S^3_3(\varepsilon,\mu) 
+\frac{5 r^4}{16 f^3}
f\left( \frac{f}{r^2} \right)'\left( \mu^2 \frac{f}{r^2} \right)' 
 S^2_3(\varepsilon,\mu) \nn && \left. 
+\frac{5 r^4}{32 f^3}f\left( \mu^2 \frac{f}{r^2} \right)'^{2}
 S^1_3(\varepsilon,\mu) 
\right\} +O\left(\frac{r^2}{L^4}\right),
      \ear
      \bear\label{B4_}
      B_4&=&\frac{1}{4\pi^2} \left\{
-\frac {\left(r^2 \right)'}{2 r^4} S^0_0(\varepsilon,\mu) 
-\frac {1}{4f}\left(\frac{f}{r^2} \right)' S^1_1(\varepsilon,\mu) 
\right. \nn && \left.-\frac {1}{4f}
\left(\mu^2\frac{f}{r^2} \right)' S^0_1(\varepsilon,\mu)\right\}
+O\left(\frac{1}{L^3}\right). 
       \ear
The mode sums and integrals in these expressions are of the form  
      \beq \label{intsums}
      S^m_n(\varepsilon,\mu)=\int \nolimits_{0}^{\infty}du 
      \cos (\varepsilon u)
      \sum \limits_{l=0}^{\infty}  \left\{  \frac{\left(l+1/2\right)^{2m+1}}
      {\left[u^2+\mu^2+\left(l+1/2\right)^2\right]^{(2n+1)/2}}-
      \mbox{subtraction terms} \right\},   
      \eeq
where $m$ and $n$ are integers, $m \geq 0$ and $n \geq -1$.
The subtraction terms for the sum over $l$  can be obtained by expanding
the function that is summed in inverse powers of $l$ and truncating at
$O(l^0)$. Such subtracting corresponds to removing the superficial
divergences in the sums over $l$ discussed above:
      \beq 
      S^1_0(\varepsilon,\mu)=\int \nolimits_{0}^{\infty}du 
      \cos (\varepsilon u)
      \sum \limits_{l=0}^{\infty}  \left\{  \frac{\left(l+1/2\right)^{3}}
      {\sqrt{u^2+\mu^2+\left(l+1/2\right)^2}}- \left(l+\frac12 \right)^2
      +\frac{u^2+\mu^2}{2} \right\},   
      \eeq
      \beq 
      S^0_{-1}(\varepsilon,\mu)=\int \nolimits_{0}^{\infty}du 
      \cos (\varepsilon u)
      \sum \limits_{l=0}^{\infty}  \left\{  \left(l+\frac12 \right)
      \sqrt{u^2+\mu^2+\left(l+1/2\right)^2}- \left(l+\frac12 \right)^2
      -\frac{u^2+\mu^2}{2} \right\},   
      \eeq
      \beq 
      S^m_m(\varepsilon,\mu)=\int \nolimits_{0}^{\infty}du 
      \cos (\varepsilon u)
      \sum \limits_{l=0}^{\infty}  \left\{  \frac{\left(l+1/2\right)^{2m+1}}
      {\left[u^2+\mu^2+\left(l+1/2\right)^2\right]^{(2m+1)/2}}- 1 \right\}.   
      \eeq
For other quantities $S^m_n(\varepsilon,\mu)$ there are no subtraction terms.
The details of calculations of $S^m_n(\varepsilon,\mu)$ in the limit 
$\varepsilon \rightarrow 0$  are discussed in Appendix A:
     \beq\label{start}
     S^1_0(\varepsilon, \mu)=\frac{4}{\varepsilon^4}-\frac{\mu^2}
     {\varepsilon^2}+\left( \frac{7}{960}-\frac{\mu^4}{4} \right)
     \left( C+\frac12\ln\left| \frac{\mu^2 \varepsilon^2}{4} \right|\right)
     +\frac{3}{16}\mu^4+\mu^4 I_2(\mu)  
     +O\left(\varepsilon^2 \ln \left|\varepsilon \right| \right),
     \eeq
     \bear
     S^{0}_{-1}(\varepsilon, \mu)&=&-\frac{2}{\varepsilon^4}-\frac{1}
     {\varepsilon^2}\left( \frac{\mu^2}{2}-\frac{1}{24} \right)
     +\left( \frac{\mu^4}{8}-\frac{\mu^2}{48}+ \frac{7}{1920} \right)
     \left( C+\frac12\ln\left| \frac{\mu^2 \varepsilon^2}{4} \right|\right)
     -\frac{3}{32}\mu^4\nn &&+\frac{\mu^2}{96}-\frac{7}{3840}
     -\frac{\mu^4}{2} I_1(\mu)+\frac{\mu^4}{2} I_2(\mu)  
     +O\left(\varepsilon^2 \ln \left|\varepsilon \right| \right),
     \ear
     \bear
     S^{0}_{0}(\varepsilon, \mu)&=&\frac{1}{\varepsilon^2}
     +\left( \frac{\mu^2}{2}-\frac{1}{24} \right)
     \left( C+\frac12\ln\left| \frac{\mu^2 \varepsilon^2}{4} \right|\right)
     -\frac{\mu^2}{4}-\mu^2 I_1(\mu)+\varepsilon^2 \left[
      -\frac{5}{64}\mu^4 +\frac{\mu^2}{96}\right.\nn
      &&\left.-\frac{7}{3840}+\left(\frac{\mu^4}{16}
     -\frac{\mu^2}{96}
     +\frac{7}{3840}\right)\left( C+\frac12\ln\left| \frac{\mu^2
      \varepsilon^2}{4} \right|\right) 
      -\frac{\mu^4}{4} I_1(\mu)+\frac{\mu^4}{4} I_2(\mu)\right]  \nn &&
     +O\left(\varepsilon^4 \ln \left|\varepsilon \right| \right),
     \ear
     \bear
     S^{n}_{n}(\varepsilon, \mu)&=&\left(\sum \nolimits_{k=0}^{n} 
     \frac{(-1)^{n+k}n!}{k!(n-k)!(2n-2k-1)}\right)\left[
     -\frac{1}{\varepsilon^2}
     - \frac{\mu^2}{2}\left( C+\frac12\ln\left| \frac{\mu^2 \varepsilon^2}
     {4} \right|\right)+\frac{\mu^2}{4}\right]\nn
     &&-\frac{\mu^2}{(2n-1)!!}
     \left( \frac{d}{\mu d \mu} \right)^n \left( \mu^{2n} I_1(\mu) \right) 
     +O\left(\varepsilon^2 \ln \left|\varepsilon \right| \right) 
     \quad (n\geq 1),
     \ear
     \bear
     S^{0}_{1}(\varepsilon, \mu)&=&-\left( C+\frac12\ln\left| 
     \frac{\mu^2 \varepsilon^2}{4} \right|\right)+
     \mu \frac{d}{d \mu}I_0(\mu)
     +\varepsilon^2\left[\left(-\frac{\mu^2}{4}+\frac{1}{48}\right)
     \left( C+\frac12\ln\left| \frac{\mu^2\varepsilon^2}{4} \right|
     \right)\right.\nn
     &&\left.+\frac{\mu^2}{4}-\frac{1}{96}+\frac{\mu^2}{2}I_1(\mu)
     \right]+O\left(\varepsilon^4 \ln \left|\varepsilon \right| \right),
     \ear
     \bear
     S^{n}_{n+1}(\varepsilon, \mu)&=&\frac{1}{(2n+1)!!}\left\{-2^n n!
     \left( C+\frac12\ln\left|\frac{\mu^2 \varepsilon^2}{4} \right|\right)
     +\mu^2\left( \frac{\partial}{\mu \partial\mu} \right)^{n+1}
     \left(\mu^{2n}I_0(\mu)\right) \right.\nn
     &&\left.+\varepsilon^2\left[
    -\mu^2 2^{n-2}n!
     \left( C+\frac12\ln\left| \frac{\mu^2\varepsilon^2}{4} \right|\right)
     +\mu^2 2^{n-2}n!+\frac{\mu^2}{2}\left( \frac{\partial}
     {\mu \partial \mu} \right)^{n}\left(\mu^{2n}I_1(\mu)\right)\right] 
     \right\}\nn &&
   +O\left(\varepsilon^4 \ln \left|\varepsilon \right| \right) \quad (n\geq 1),
     \ear
     \bear\label{finish}
     S^{m}_{n}(\varepsilon, \mu)&=&\frac{(2m+1)!!}{(2n+1)!!}\left(
     -\frac{\partial}{\mu \partial \mu} \right)^{n-m-1} S^{m}_{m+1}
     (\varepsilon, \mu) \quad (m\geq 0, \ n\geq m+2),
     \ear
where
     \beq
     I_n(\mu)=\int \nolimits_{0}^{\infty}\frac{x^{2n-1}\ln|1-x^2|}
     {1+e^{2\pi \mu x}}dx.
     \eeq
Substitution of these expressions into Eqs. (\ref{B1})-(\ref{B4})
and then into Eqs. (\ref{Ttt})-(\ref{Tthth}) gives nontrivial components of 
$\left< T^{\mu}_{\nu} \right>_{unren}$.

The renormalization of $\left< T^{\mu}_{\nu} \right>$ is achieved by
subtracting from $\left< T^{\mu}_{\nu} \right>_{unren}$ 
the counterterms $\left< T^{\mu}_{\nu} \right>_{DS} $
 and then letting $\tilde \tau \rightarrow \tau$:
      \beq
      \left< T^{\mu}_{\nu} \right>_{ren}=\lim_{\tilde \tau \rightarrow \tau}
      \left[\left< T^{\mu}_{\nu} \right>_{unren}-
      \left< T^{\mu}_{\nu} \right>_{DS}\right].
      \eeq
The expression for $\left< T^{\mu}_{\nu} \right>_{DS}$ 
are displayed in Ref. \cite{Chris} and are determined by $\sigma$, one 
half the square of the distance between the points $x$ and $\tilde x$
along the shortest geodesic connecting them, and 
$\sigma^{\mu}$, covariant derivative of $\sigma$. 
For the metric (\ref{metric}) the calculations of these quantities
 and $g^{\mu \tilde \nu}$ are similar to those in Ref. \cite{AHS}:
      \bear
      \sigma^t&=&(t-\tilde t)+\frac{f'^2}{24f}(t-\tilde t)^3
      +O\left((t-\tilde t)^5\right),\nn 
      \sigma^{\rho}&=&-\frac{f'}{4}(t-\tilde t)^2
      +O\left((t-\tilde t)^4\right),\nn 
      \sigma^{\theta}&=&\sigma^{\phi}=0, \nn
      \sigma&=&\frac12 g_{\mu \nu}\sigma^{\mu}\sigma^{\nu},
      \ear
      \bear
    g^{t \tilde t}&=&-\frac{g^{\rho \tilde \rho}}{f}=-\frac1f-\frac{f'^2}{8f^2}
     (t-\tilde t)^2+O\left((t-\tilde t)^4\right),\nn 
      g^{t \tilde \rho}&=&-g^{\rho \tilde t}=-\frac{f'}{2f}(t-\tilde t)
      +O\left((t-\tilde t)^3\right) 
      \ear
and  
\bear
\left< T^{t}_{t} \right>_{DS} &=&
\frac{1}{4 \pi^2}\left(
-{\frac {6}{ r^4 \varepsilon^4}}
-\frac{1}{r^2{\varepsilon}^2}\left[
-{\frac {{m}^{2}}{2}}
+{\frac {1}{6r^2}}
-{\frac {\xi}{r^2}}
+\left (
{\frac {2\xi\,f''}{f}}
+{\frac {(r^2)'^2}{24r^4}}
-{\frac {(r^2)''}{6r^2}}
+{\frac {2\xi f' (r^2)'}{r^2f}} \right. \right. \right. \nn && \left. \left.
-{\frac {\xi\,(r^2)'^2}{4r^4 }}
+{\frac {\xi\,(r^2)''}{ r^2 }}
-{\frac {f'(r^2)'}{3fr^2}} 
-{\frac {f''}{3f}}
-{\frac {f'^2}{12 f^2}}
-{\frac {\xi\,f'^2}{f^2}}
\right )
\right]                                                            
+\left\{
-{\frac {{m}^{4}}{8}}  
+\left ({\frac {1}{12r^2}}                 \right. \right. \nn && \left.
-{\frac {\xi}{2r^2}}\right ){m}^{2} 
-{\frac {{\xi}^{2}}{2r^4}}
+{\frac {\xi}{6r^4}}
-{\frac {1}{60r^4}}
+\left[
\left (-{\frac {(r^2)''}{12r^2}}
+{\frac {\xi\,(r^2)''}{2 r^2}}
-{\frac {\xi\,(r^2)'^2}{8 r^4}}
+{\frac {(r^2)'^2}{48r^4}}\right ){m}^{2}      \right. \nn && \left. \left.
+{\frac {3{\xi}^{2}(r^2)'^2}{4r^6}}
-{\frac {\xi\,(r^2)'^2}{4r^6}}
+ {\frac {(r^2)'^{2}}{48 r^6}}
\right]
\right\}{\left( C+\frac12 \ln \left| \frac{\mds^2 r^2  \varepsilon^2}{4 }
\right| \right)}
-{\frac {{m}^{4}}{32}}
+\left (
{\frac {1}{24 r^2}}        \right. \nn && \left.
-{\frac {\xi}{4 r^2}}
\right ){m}^{2}                                   
-{\frac {{\xi}^{2}}{2r^4}}
+{\frac {\xi}{6 r^4}}
-{\frac {1}{60 r^4}}
+\left[
\left (
{\frac {(r^2)'^2}{96 r^4}}
-{\frac {f''}{24 f}}
-{\frac {f'^2}{32f^2}}
-{\frac {\xi\,f'^2}{4 f^2}}
+{\frac {\xi f'(r^2)'}{2f r^2}}  \right. \right. \nn && \left. \left.
+{\frac {\xi\,(r^2)''}{4 r^2}}
-{\frac {\xi (r^2)'^{2}}{16 r^4}}
-{\frac {f'(r^2)'}{24 r^2 f}} 
+{\frac {\xi\,f''}{2 f}}
-{\frac {(r^2)''}{24 r^2 }}
\right ){m}^{2}                          
-{\frac {{\xi}^{2}f'^2}{2 r^2 f^2}}
-{\frac {(r^2)'^2}{48 r^6}}
-{\frac {\xi\,f''}{4 r^2 f}}      \right. \nn && \left.
+{\frac {{\xi}^{2}f''}{ r^2 f }}
-{\frac {{\xi}^{2}(r^2)'^{2}}{4 r^6}}
-{\frac {5\xi\,(r^2)''}{12 r^4}} 
+{\frac {f'^2}{96 r^2 f^2}}
+{\frac {{\xi}^{2}(r^2)''}{r^4}}
-{\frac {f'\xi\,(r^2)'}{6r^4 f}}
+{\frac {f'{\xi}^{2}(r^2)'}{r^4 f}}           \right. \nn && \left. \left.
+{\frac {\xi\,f'^2}{48 r^2 f^2}}
+{\frac {f''}{72 r^2f}}
+{\frac {(r^2)''}{24 r^4}}
+{\frac {\xi\,(r^2)'^{2}}{6 r^6}}
\right] \right)+O\left(\frac{1}{L^4}\right),
\ear
\bear
\left< T^{\rho}_{\rho} \right>_{DS} &=&\frac1{4 \pi^2}
\left(
\frac {2}{r^4 \varepsilon^4}
-\frac1{r^2 \varepsilon^2}\left[
{\frac {{m}^{2}}{2}}
+{\frac {\xi}{r^2}}
-{\frac {1}{6r^2}}
+\left ({\frac {5f'^{2}}{12 f^{2}}}
-{\frac {\xi\,(r^2)'^2}{4r^4}}
+{\frac {(r^2)'^2}{24r^4}}
-{\frac {f'(r^2)'}{12 fr^2}} \right. \right. \right. \nn && \left. \left.
+{\frac {\xi\,f'(r^2)'}{2 fr^2}}
\right )
\right]
+\left\{
-\frac{m^4}{8}
+\left (
-{\frac {\xi}{2r^2}}
+\frac1{12r^2}\right ){m}^{2}
-{\frac {{\xi}^{2}}{2r^4}}
+{\frac {\xi}{6r^4}}
-{\frac {1}{60r^4}}                            \right. \nn && \left.          
+\left [ \left (
{\frac {\xi\,f'(r^2)'}{4fr^2}}
-{\frac {(r^2)'^2}{48r^4}}   
+{\frac {\xi\,(r^2)'^2}{8r^4}}
-{\frac {f'(r^2)'}{24r^2f}}\right ){m}^{2}
+{\frac {\xi\,(r^2)'^2}{4r^6}}
-{\frac {3{\xi}^{2}(r^2)'^2}{4r^6}}  \right. \right. \nn && \left. \left. 
-{\frac {(r^2)'^2}{48r^6}}\right] 
\right\}\left( C+\frac12 \ln \left| \frac{\mds^2 r^2  \varepsilon^2}{4 }
\right| \right) 
+{\frac {3{m}^{4}}{32}}
+\left (
{\frac {\xi}{4r^2}} 
-\frac1{24r^2}
\right ){m}^{2} 
+\left[
\left(
{\frac {(r^2)'^2}{96r^4}} \right. \right. \nn && \left. \left.
-{\frac {\xi\,(r^2)'^2}{16r^4}}
+{\frac {\xi\,f'(r^2)'}{8fr^2}}
-{\frac {f'(r^2)'}{48fr^2}}
+{\frac {5f'^{2}}{96f^{2}}}\right ){m}^{2}
-{\frac {5f'^{2}}{288f^{2}r^2}}
+{\frac {f'(r^2)'}{48fr^4}}     \right. \nn && \left. \left.
-{\frac {5\xi\,f'(r^2)'}{24fr^4}}
+{\frac {5\xi\,f'^{2}}{48f^{2}r^2}}
+{\frac {f'{\xi}^{2}(r^2)'}{2fr^4}}
\right] \right)+O\left(\frac{1}{L^4}\right),
\ear
\bear
\left< T^{\theta}_{\theta} \right>_{DS} &=&
\frac{1}{4 \pi^2}\left(
{\frac {2}{r^4 \varepsilon^4 }}
-\frac{1}{r^2 \varepsilon^2}\left[
{\frac {{m}^{2}}{2}}
+\left (
{\frac {\xi\,(r^2)'^{2}}{4r^4}}
+{\frac {(r^2)''}{12r^2}}
-{\frac {(r^2)'^{2}}{24r^4}}
-{\frac {f''}{12f}}
-{\frac {\xi\,f'^{2}}{4f^2}}
-{\frac {f'^{2}}{24f^2}}     \right. \right. \right. \nn && \left. \left.
+{\frac {\xi\,f'(r^2)'}{4r^2f}}
-{\frac {(r^2)'f'}{24r^2f}}
-{\frac {\xi\,(r^2)''}{2r^2}}
+{\frac {\xi\,f''}{2f}}
\right )     
\right] 
+\left\{
-\frac{m^4}{8}
+{\frac {{\xi}^{2}}{2r^4}}
-{\frac {\xi}{6r^4}}
+{\frac {1}{60r^4}}         \right. \nn &&  
+\left [  
\left (
{\frac {\xi\,f''}{4f}} 
-{\frac {\xi\,(r^2)'^{2}}{8r^4}}
+{\frac {(r^2)'^{2}}{48r^4}}
-{\frac {(r^2)''}{24r^2}} 
-{\frac {\xi\,f'^{2}}{8f^2}}
+{\frac {f'^{2}}{48f^2}}  
+{\frac {\xi\,(r^2)'f'}{8r^2f}}
-{\frac {(r^2)'f'}{48r^2f}} \right. \right.  \nn &&  \left.
-{\frac {f''}{24f}}
+{\frac {\xi\,(r^2)''}{4r^2}}
\right ){m}^{2} 
-{\frac {(r^2)''}{24r^4}}
+{\frac {(r^2)'^{2}}{24r^6}}
-{\frac {(r^2)'f'}{48r^4f}}
-{\frac {3{\xi}^{2}f'(r^2)'}{4fr^4}}
-{\frac {\xi\,(r^2)'^{2}}{2r^6}}  \nn &&   \left. \left. 
+{\frac {\xi\,(r^2)''}{2r^4}}
+{\frac {\xi\,(r^2)'f'}{4r^4f}}
+{\frac {3{\xi}^{2}(r^2)'^{2}}{2r^6}}
-{\frac {3{\xi}^{2}(r^2)''}{2r^4}}
\right ]
\right\}{\left( C+\frac12 \ln \left| \frac{m^2 r^2  \varepsilon^2}{4 } 
\right| \right)}
+{\frac {3m^4}{32}}    \nn && 
+\left [
\left (
{\frac {\xi\,(r^2)'f'}{16r^2f}}
+{\frac {\xi\,(r^2)'^{2}}{16r^4}}
-{\frac {\xi\,f'^{2}}{16f^2}}
-{\frac {(r^2)'^{2}}{96r^4}}
-{\frac {\xi\,(r^2)''}{8r^2}}
+{\frac {\xi\,f''}{8f}}  
+{\frac {(r^2)''}{48r^2}}
-{\frac {(r^2)'f'}{96r^2f}}                 \right.  \right.  \nn && \left. 
-{\frac {f''}{48f}}
\right ){m}^{2}           
+{\frac {{\xi}^{2}f''}{2r^2f}} 
+{\frac {5\xi\,f'^{2}}{48r^2f^2}}
+{\frac {f''}{48r^2f}}
-{\frac {\xi\,f'(r^2)'}{6r^4f}}
+{\frac {f'(r^2)'}{48r^4f}}
-{\frac {5\xi\,f''}{24r^2f}}
-{\frac {{\xi}^{2}f'^{2}}{4r^2f^2}}                   \nn && \left. \left.
+{\frac {f'{\xi}^{2}(r^2)'}{4r^4f}}
-{\frac {f'^{2}}{96r^2f^2}}
\right ]  \right )
+O\left(\frac{1}{L^4}\right),             
\ear
where $C$ is Euler's constant and $\mds$ is equal to the
mass $m$ of the field for a massive scalar field. For a massless scalar field
it is an arbitrary parameter due to the infrared cutoff in
$\left< T^{\mu}_{\nu} \right>_{DS}$. A particular choice of the value of 
$\mds$ orresponds to a finite renormalization of the coefficients of
terms in the gravitational Lagrangian and must be fixed by experiment or 
observation.

The procedure described above gives a second-order WKB 
approximation of $\left< T^{\mu}_{\nu} \right>_{ren}$. The zeroth-order
expressions of nontrivial components  of
$\left< T^{\mu}_{\nu} \right>_{ren}$ have the form
     \bear \label{T_0_tt}
     \left< T^{t}_{t} \right>_{ren}=\left< T^{\rho}_{\rho} \right>_{ren}&=&
     \frac{1}{4 \pi^2}\left\{\frac{m^2}{8 r^2}\left( \xi-\frac18 \right)  
     +\frac1{r^4}\left( \frac{79}{7680}-\frac{11}{96}\xi+\frac38 \xi^2 \right)
   \right.\nn &&+ \left[-\frac{m^4}8+\frac{m^2}{2 r^2}\left(\frac16-\xi \right)
     +\frac1{r^4}\left( -\frac{1}{60}+\frac{1}{6}\xi-\frac12 \xi^2 \right)
      \right]\ln\sqrt{\frac{\mu^2}{\mds^2r^2}}\nn &&\left.
     + \left[\frac{m^4}2+\frac{2m^2}{r^2}\left(\xi-\frac18 \right)  
     +\frac2{r^4}\left( \xi-\frac18 \right)^2 \right] 
     \left[I_1(\mu)-I_2(\mu)\right]  \right\},
     \ear
     \bear \label{T_0_thth}
     \left< T^{\theta}_{\theta} \right>_{ren}=
     \left< T^{\varphi }_{\varphi} \right>_{ren}&=&
     \frac{1}{4 \pi^2}\left\{\frac{m^2}{8 r^2}\left( \xi-\frac18 \right)  
     -\frac1{8 r^4} (\xi-\frac18)^2+ \left[-\frac{m^4}8  
     +\frac1{r^4}\left(\frac{1}{60}-\frac{1}{6}\xi
      \right.\right.\right.\nn &&\left.\left.+\frac12 \xi^2 \right)
      \right]\ln\sqrt{\frac{\mu^2}{\mds^2r^2}}
     + \left[\frac{m^2}{ r^2}\left(\frac18-\xi \right)  
     -\frac2{r^4}\left( \xi-\frac18 \right)^2 \right]I_1(\mu)
      \nn && \left.+ \left[\frac{m^4}{2} 
     +\frac{2m^2}{r^2}\left(\xi-\frac18 \right) 
      +\frac2{r^4}\left( \xi-\frac18 \right)^2 
      \right] I_2(\mu)  \right\}.
     \ear
The second-order expressions of $\left< T^{\mu}_{\nu} \right>_{ren}$ are 
given in Appendix B.

Let us stress that the renormalized expectation value of the stress-energy
tensor (\ref{T_0_tt}), (\ref{T_0_thth}) are {\em exact} if $f(\rho)=f_0=const$
and $r(\rho)=r_0=const$, i.e., in the spacetime with metric
        \beq
        ds^2=-f_0dt^2+d\rho^2+r_0^2(d\theta^2+\sin^2\theta\, d\varphi^2). 
        \eeq
The expressions (\ref{T_0_tt}) and (\ref{T_0_thth}) may be simplified
in the cases $L^2 \gg r^2, \; m^2 r^2 \ll 1$ and 
$L^2 \gg r^2, \; m^2 r^2 \gg 1$.

%
\subsection{The case $L^2 \gg r^2, \; m^2 r^2 \ll 1$}
%
					
In this case
      \bear \label{ll1}
      \left< T^{t}_{t} \right>_{ren}=\left< T^{\rho}_{\rho} \right>_{ren}&=&
     \frac{1}{4 \pi^2 r^4}\left\{ 
     \left( \frac38 \xi^2-\frac{11}{96}\xi +\frac{79}{7680} \right) 
     +\left( -\frac12 \xi^2+\frac{1}{6}\xi -\frac{1}{60} \right)\ln\sqrt
     {\frac{8\xi-1}{4\mds^2r^2}}
     \right.\nn &&\left.+\left( 2 \xi^2-\frac{1}{2}\xi+\frac{1}{32} 
     \right)\left[
     I_1\left( 2 \xi-\frac14 \right)-I_2\left( 2 \xi-\frac14 \right) \right]
     \right\}\nn &&+O\left( \frac{m^2}{r^2} \right)+O\left( \frac1{L^2r^2} 
     \right),
      \ear
      \bear \label{ll2}
      \left< T^{\theta}_{\theta} \right>_{ren}=
      \left< T^{\varphi }_{\varphi} \right>_{ren}&=&
     \frac{1}{4 \pi^2 r^4}\left\{ 
     \left(-\frac18 \xi^2+\frac{1}{32}\xi-\frac{1}{512} \right) 
     +\left(\frac12 \xi^2-\frac{1}{6}\xi +\frac{1}{60} \right)
      \ln\sqrt{\frac{8\xi-1}{4\mds^2r^2}}
     \right.\nn &&\left.+\left(-2 \xi^2+\frac{1}{2}\xi-\frac{1}{32} \right) 
     \left[
     I_1\left( 2 \xi-\frac14 \right)-I_2\left( 2 \xi-\frac14 \right) \right]
     \right\} \nn &&+O\left( \frac{m^2}{r^2} \right)+O\left( \frac1{L^2r^2} 
     \right).
      \ear
In particular, if $\xi=1/6$,
      \beq
      I_1(\sqrt{3}/6)\approx-0.05962, \quad I_2(\sqrt{3}/6)\approx0.50385
      \eeq
and
\bear
   \left< T^{t}_{t} \right>_{ren}=\left< T^{\rho}_{\rho} \right>_{ren}&\approx&
     \frac{1}{4 \pi^2 r^4}\left[0.00310 
     -\frac1{360} \ln\frac{1}{\mds\left| r \right|}
      \right]+O\left( \frac{m^2}{r^2} \right)+O\left( \frac1{L^2r^2} \right),
      \ear
      \bear
      \left< T^{\theta}_{\theta} \right>_{ren}&\approx&
     \frac{1}{4 \pi^2 r^4}\left[ -0.00171
     +\frac1{360}\ln\frac{1}{\mds\left|r\right|}
      \right]+O\left( \frac{m^2}{r^2} \right)+O\left( \frac1{L^2r^2} \right).
      \ear

%
\subsection{The case $L^2 \gg r^2, \;  m^2 r^2 \gg 1$}
%
					
In this case $\mds=m$. Therefore
      \beq
      \ln\sqrt{\frac{\mu^2}{m^2r^2}}=\frac{8\xi-1}{8m^2r^2}
      -\frac{(8\xi-1)^2}{64m^4r^4}+\frac{(8\xi-1)^3}{384m^6r^6}
      +O\left(\frac1{m^8r^8}\right),
      \eeq
       \bear
       I_1(\mu)&=&-\frac7{1920\; m^4r^4}+\frac{1}{m^6 r^6}
      \left[ \frac{7}{480}\left(\xi-\frac1{8}\right)-\frac{31}{32256} \right] 
        +O\left(\frac1{m^8r^8}\right), \nn 
        I_2(\mu)&=&-\frac{31}{16128\; m^6r^6}+O\left(\frac1{m^8r^8}\right)
       \ear
and
       \bear \label{gg}
       \left< T^{\mu}_{\nu} \right>_{ren}= \left( \begin{array}{rrrr}
        -1& 0 & 0 & 0 \\ 0 & \quad -1& 0& 0 \\ 0 & 0 & \quad 2& 0\\ 0 &0 & 0& \quad 2\end{array} 
       \right) 
       \left[\frac1{4\pi^2m^2r^6} \right. && \left.\left( \frac{\xi^3}{6}-\frac{\xi^2}{12}
       +\frac{\xi}{60}
       -\frac1{630} \right)
       \right. \nn && \left.+O\left(\frac1{m^4 r^8} \right)
       +O\left( \frac1{m^2L^2r^4} \right) \right].
        \ear  
As was demonstrated by Morris and Thorne \cite{MT} the radial tension 
$\tau$ at the wormhole throat  must be positive. In the considered case
$\tau=- \left< T^{\rho}_{\rho} \right>_{ren} >0$ for
       \beq
        \xi>\frac{\left(9800+735\sqrt{205}\right)^{1/3}}{210}
       - \frac7{6(9800+735\sqrt{205})^{1/3}}+\frac16 \approx 0.2538.
        \eeq

As an example of a metric function $r^2(\rho)$ satisfying the condition
$L^2 \gg r^2, \;  m^2 r^2 \gg 1$ one can consider a function growing as 
a logarithm: 
        \beq \label{ln}
        ds^2=-dt^2+d\rho^2+r_0^2\left[ 1+\alpha^2\ln \left( 1
        +\frac{\rho^2}{\rho_0^2} \right)  \right]\left( d \theta^2 
        +\sin ^2\theta d \phi^2\right), 
        \eeq
where $r_0, \alpha$, and $\rho_0$ are constants. The conditions 
$L^2 \gg r^2$ and $r^2 m^2 \gg 1$ gives 
$\alpha^2 r_0^2/\rho_0^2 \ll 1$ and $r_0^2 m^2 \gg 1$, respectively.


\section{Conclusion}

 
We have obtained an analytical approximation of the stress-energy 
tensor of quantized scalar fields in static spherically symmetric 
spacetimes with topology $S^2 \times R^2$.  For some values 
of coupling to the scalar curvature $\xi$
the stress-energy tensor obtained has the needed exotic (in the sense 
of Morris and Thorne \cite{MT}) properties to support a static wormhole.
All three approximations (\ref{T_0_tt}), (\ref{T_0_thth}), (\ref{ll1}),  
(\ref{ll2}) and 
(\ref{gg}) for $\left< T^{\mu}_{\nu} \right>_{ren}$ are conserved, i. e., 
$$\left< T^{\mu}_{\nu} \right>_{ren \, ; \mu }=0,$$
and for the conformally invariant field the approximation for 
$\left< T^{\mu}_{\nu} \right>_{ren}$  has a trace equal to 
the trace anomaly. These approximations
are determined by the correlation between three length scales: 
$L$ (\ref{Lm}), $r$ (\ref{metric}) and the Compton length 
$1/m$. The use of the expression obtained for
$\left< T^{\mu}_{\nu} \right>_{ren}$ as a source term in the semiclassical
Einstein field equations 
$G^{\mu}_{\nu}=8 \pi G \left< T^{\mu}_{\nu} \right>_{ren}$  
demands some accuracy. First of all
in these equations a new length scale $\sqrt{G}$ appears. 
If on the left-hand side of these equations we neglect the terms of order 
$1/L^2$, the system of equations will be algebraically incompatible.
If we keep such terms the terms of corresponding order 
should be kept on the right-hand side of the equations.

In conclusion let us note that in the case $L^2 \lsim r^2$ the WKB parameter 
$\lambda_{WKB}$ [Eq. (\ref{lwkb})] coinsides with the small parameter $1/(mL)$
of the DeWitt-Schwinger expansion of $\left< T^{\mu}_{\nu} \right>$
and with the small parameter used in \cite{AHS} to obtain an 
analytical approximation for $\left< T^{\mu}_{\nu} \right>_{ren}$. 
In this case as the first and second  WKB orders of 
$\left< T^{\mu}_{\nu} \right>_{ren}$ vanish 
[see Eqs. (\ref{wkb2_tt})-(\ref{wkb2_thth})]
and the next orders are needed for calculation of these quantities.

\section*{Acknowledgment}
This work was supported by the Russian Foundation for Basic
Research Grant No. 99-02-17941. 
  
\section*{Appendix A}

\setcounter{equation}{0}
\renewcommand{\theequation}{A\arabic{equation}}

The sums over $l$ in expression (\ref{intsums}) can be evaluated by 
using the following method:
      \beq
      J_n=2i\sum \limits_{l=0}^{n} f( l+1/2) =
      -2\pi i \sum \limits_{l=0}^{n} \mathrel{\mathop{res}\limits_{z=l+1/2}} 
      \left[ f(z) \tan (\pi z) \right] = -\int \limits_{\cal C}f(z) 
      \tan (\pi z) dz,   
      \eeq
where $\cal C$ is a closed contour which surrounds a region on the
complex plane containing the poles of $\tan (\pi z)$: 
$z=l+1/2$ for $l=0, 1, 2, ..., n$ . It is 
supposed that countour $\cal C$ is passed anticlockwise, $f(z)$ 
is a holomorphic function inside this contour, and 
      \beq \label{1}
      \left| f(x+iy) \right|<\epsilon(x)e^{a|y|}, \quad 0<a<2\pi, 
      \eeq
where
      \beq
      \epsilon(x)\rightarrow 0 \quad \mbox{for} \quad x \rightarrow +\infty.
      \eeq
Let us choose the contour so that
      \bear
     J_n &=&\left\{\int \nolimits_{q}^{q-ih}+\int \nolimits_{q-ih}^{q+n+1-ih}+
      \int \nolimits_{q+n+1-ih}^{q+n+1}\right\}f(z)\tan(\pi z)dz\nn
      &&-\left\{ \int \nolimits_{q}^{q+ih}+\int \nolimits_{q+ih}^{q+n+1+ih}+
      \int \nolimits_{q+n+1+ih}^{q+n+1}\right \}f(z)\tan(\pi z)dz \nn
      &=&2i \int \nolimits_{q}^{q+n+1}f(x)dx
      -\int \nolimits_{q}^{q-ih}[f(z)-f(z+n+1)][\tan (\pi z)+i]dz \nn &&
      -\int \nolimits_{q-ih}^{q+n+1-ih} f(z)[\tan(\pi z +i)] dz 
      +\int \nolimits_{q}^{q+ih} [f(z)-f(z+n+1)] [\tan(\pi z)-i]dz \nn &&
      -\int \nolimits_{q+ih}^{q+n+1+ih}f(z)[\tan(\pi z)-i]dz,
      \ear
where $-1/2<q<1/2$ and $h>0$. However,
      \bear
     J_-=\left|\int \nolimits_{q-ih}^{q+n+1-ih} f(z)[\tan(\pi z +i)] dz\right|
      &\leq& (n+1) \max \limits_{q \leq x \leq q+n+1} 
      \left| f(x-ih) \right| \left| \tan[\pi(x-ih)] +i \right|\nn
      &&<(n+1) \max \limits_{q \leq x \leq q+n+1} 
      \epsilon(x) \frac{2 e^{ah}}{\displaystyle (e^{2\pi h}-1)}, \nn
      J_+=\left| \int \nolimits_{q+ih}^{q+n+1+ih}f(z)[\tan(\pi z)-i]dz \right|
      & \leq &(n+1)\max \limits_{q \leq x \leq q+n+1}
      \left| f(x+ih) \right| \left| \tan[\pi(x+ih)] -i \right| \nn
      &&<(n+1) \max \limits_{q \leq x \leq q+n+1} 
      \epsilon(x) \frac{2e^{ah}}{\displaystyle (e^{2\pi h}-1)},
      \ear
therefore  $J_{-}\rightarrow 0, J_{+}\rightarrow 0$ if $h\rightarrow \infty$, 
and in this case
      \bear
      J_n&=&2i \int \nolimits_{q}^{q+n+1}f(x)dx
      -\int \nolimits_{q}^{q-i\infty}[f(z)-f(z+n+1)][\tan (\pi z)+i]dz \nn
      &&+\int \nolimits_{q}^{q+i\infty} [f(z)-f(z+n+1)] [\tan(\pi z)-i]dz, 
      \ear
If we let $n\rightarrow \infty$ then
      \bear &&
      \left|\int\nolimits_{q}^{q-i\infty}f(z+n+1)[\tan(\pi z)+i]dz\right| 
      \leq \epsilon(q+n+1)\int \nolimits_{0}^{\infty}
      e^{ay}\left| \frac{2e^{-i\pi q}}{e^{-i\pi q}
      +e^{2\pi y}e^{i\pi q}} \right| dy \nn &&
      \leq const \times \epsilon(q+n+1)\int \nolimits_{0}^{\infty}
      e^{(a-2\pi)y} dy\rightarrow 0 \qquad (n \rightarrow \infty), 
      \ear
      \bear &&
      \left| \int \nolimits_{q}^{q+i\infty} f(z+n+1) [\tan(\pi z)-i]dz\right|
      \leq \epsilon(q+n+1)\int \nolimits_{0}^{\infty}
      e^{ay}\left| \frac{2e^{i\pi q}}{e^{i\pi q}
      +e^{2\pi y}e^{-i\pi q}} \right| dy \nn
      &&\leq const \ \epsilon(q+n+1)\int \nolimits_{0}^{\infty}
      e^{(a-2\pi)y} dy\rightarrow 0, \qquad (n \rightarrow \infty) 
      \ear
and
      \bear
      J_{\infty}&=&2i\sum \limits_{l=0}^{\infty} f( l+1/2) 
      =2i \int \nolimits_{q}^{\infty}f(x)dx
      -\int \nolimits_{q}^{q-i\infty}f(z)[\tan (\pi z)+i]dz \nn &&
      +\int \nolimits_{q}^{q+i\infty}f(z) [\tan(\pi z)-i]dz 
       \ear
or
     \beq
     \sum \limits_{l=0}^{\infty} f( l+1/2) 
      =\int \nolimits_{q}^{\infty}f(x)dx
      +\int \nolimits_{q-i\infty}^{q}\frac{f(z)}{1+e^{i2\pi z}}dz 
      -\int \nolimits_{q}^{q+i\infty}\frac{f(z)}{1+e^{-i2\pi z}}dz,
       \eeq
where $-1/2<q<1/2$, $f(z)$ is a holomorphic function for $Re z\geq q$,
and $f(z)$ satisfies condition (\ref{1}).

Using the last expression we can calculate the sums in Eq. (\ref{intsums}):
      \bear
      &&\sum \limits_{l=0}^{\infty}  \left\{  \frac{\left(l+1/2\right)^{3}}
      {\sqrt{u^2+\mu^2+\left(l+1/2\right)^2}}- \left(l+\frac12 \right)^2
      +\frac{u^2+\mu^2}{2} \right\}=\lim _{q\rightarrow 0}\left\{
       \int\nolimits_{q}^{\infty}\left[\frac{x^3}{\sqrt{u^2+\mu^2+x^2}}
      \right. \right.\nn  && \left.-x^2 +\frac{u^2+\mu^2}{2}\right]dx 
      +\int \nolimits_{q-i\infty}^{q} \left[ \frac{z^3}
      {\sqrt{u^2+\mu^2+z^2}}-z^2+\frac{u^2+\mu^2}{2}\right]\frac{dz}
      {(1+e^{i2\pi z})} \nn && \left. -\int \nolimits_{q}^{q+i\infty}\left[ 
      \frac{z^3}
      {\sqrt{u^2+\mu^2+z^2}}-z^2+\frac{u^2+\mu^2}{2}\right]\frac{dz}
      {(1+e^{-i2\pi z})}  \right\}
      =\frac23 \left(u^2+\mu^2\right)^{3/2} \nn &&
      -2\int \nolimits_{0}^{\sqrt{u^2+\mu^2}}\frac{dx}{(1+e^{2\pi x})}
      \frac{x^3}{\sqrt{u^2+\mu^2-x^2}}, 
      \ear
      \bear
      &&\sum \limits_{l=0}^{\infty}  \left\{\left(l+\frac12\right)
      \sqrt{u^2+\mu^2+\left(l+1/2\right)^2}- \left(l+\frac12 \right)^2
      -\frac{u^2+\mu^2}{2} \right\}=\lim _{q\rightarrow 0}\left\{
      \int\nolimits_{q}^{\infty}\left[x \sqrt{u^2+\mu^2+x^2}
      \right. \right. \nn && \left. -x^2
      -\frac{u^2+\mu^2}{2}\right]dx
      +\int \nolimits_{q-i\infty}^{q} \left[z
      \sqrt{u^2+\mu^2+z^2}-z^2-\frac{u^2+\mu^2}{2}\right]\frac{dz}
      {(1+e^{i2\pi z})} \nn && \left.  -\int \nolimits_{q}^{q+i\infty}\left[z
      \sqrt{u^2+\mu^2+z^2}-z^2-\frac{u^2+\mu^2}{2}\right]\frac{dz}
      {(1+e^{-i2\pi z})}  \right\}
      =-\frac{\left(u^2+\mu^2\right)^{3/2}}{3}  \nn &&
      +2\int \nolimits_{0}^{\sqrt{u^2+\mu^2}}\frac{x\sqrt{u^2+\mu^2-x^2}}
      {(1+e^{2\pi x})}dx,
      \ear
      \bear
      &&\sum \limits_{l=0}^{\infty}  \left\{  \frac{\left(l+1/2\right)^{2m+1}}
      {\left[u^2+\mu^2+\left(l+1/2\right)^2\right]^{(2m+1)/2}}-1\right\}
     =\lim _{q\rightarrow 0}\left\{ \int \nolimits_{q}^{\infty} \left[ 
     \frac{x^{2m+1}}{\left(u^2+\mu^2+x^2\right)^{(2m+1)/2}}-1\right]dx\right.
     \nn && +\int \nolimits_{q-i\infty}^{q} \left[ \frac{z^{2m+1}}
    {\left(u^2+\mu^2+z^2\right)^{(2m+1)/2}}-1\right]\frac{dz}{(1+e^{i2\pi z})}
     -\int \nolimits_{q}^{q+i\infty} \left[ \frac{z^{2m+1}}
     {\left(u^2+\mu^2+z^2\right)^{(2m+1)/2}} \right. \nn && \left. \left.
     -1 \frac{}{} \right]\frac{dz}{(1+e^{-i2\pi z})}
      \right\}=\sum \nolimits_{k=0}^{m}\frac{(-1)^{m+k}m!}{k!(m-k)!}
      \frac{\sqrt{u^2+\mu^2}}{(2m-2k-1)}\nn
      &&+2(-1)^m\lim _{\delta\rightarrow 0}\left[\int \nolimits_{0}^{
      \sqrt{u^2+\mu^2}-\delta}\frac{x^{2m+1}}{(1+e^{2\pi x})
      \left(u^2+\mu^2-x^2\right)^{(2m+1)/2}}dx \right. \nn && \left.
      -\left(\mbox{terms of this integral that}\atop\mbox
      { diverge in the limit}\ \delta \rightarrow 0  \right) \right]
      =\sum \nolimits_{k=0}^{m}\frac{(-1)^{m+k}m!}{k!(m-k)!}
      \frac{\sqrt{u^2+\mu^2}}{(2m-2k-1)}
      \nn && +\frac{2}{(2m-1)!!}
      \int \nolimits_{0}^{\sqrt{u^2+\mu^2}}\frac{xdx}{\sqrt{u^2+\mu^2-x^2}}
      \left( \frac{d}{xdx} \right)^m \frac{x^{2m}}{(1+e^{2\pi x})}
      \quad (m \geq 0),
      \ear
      \bear
      &&\sum \limits_{l=0}^{\infty} \frac{\left(l+1/2\right)^{2m+1}}
      {\left[u^2+\mu^2+\left(l+1/2\right)^2\right]^{(2m+3)/2}} 
      =\lim _{q\rightarrow 0}\left\{
      \int\nolimits_{q}^{\infty}\frac{x^{2m+1}}{\left(u^2+\mu^2+x^2
      \right)^{(2m+3)/2}}dx\right.\nn
      &&\left.+\int \nolimits_{q-i\infty}^{q}  \frac{z^{2m+1}}
      {(u^2+\mu^2+z^2)^{(2m+3)/2}}\frac{dz}
      {(1+e^{i2\pi z})}-\int \nolimits_{q}^{q+i\infty}\frac{z^{2m+1}}
      {(u^2+\mu^2+z^2)^{(2m+3)/2}}\frac{dz}{(1+e^{-i2\pi z})}\right\}\nn
      &&=\frac{2^m m!}{(2m+1)!!\sqrt{u^2+\mu^2}}
      +2(-1)^m\lim _{\delta\rightarrow 0}\left[\int 
      \nolimits_{0}^{\sqrt{u^2+\mu^2}-\delta}\frac{x^{2m+1}}
      {(u^2+\mu^2-x^2)^{(2m+3)/2}}\frac{dx}{(1+e^{2\pi x})}\right.\nn
      &&\left.-\left(\mbox{terms of this integral that}\atop\mbox
      { diverge in the limit}\ \delta \rightarrow 0  \right) \right] \nn &&
      =-\frac{2}{(2m+1)!!}\int \nolimits_{0}^{\sqrt{u^2+\mu^2}} 
      \frac{xdx}{\sqrt{u^2+\mu^2-x^2}}\left( \frac{d}{xdx} \right)^{m+1}
      \frac{x^{2m}}{1+e^{2\pi x}} 
      \quad (m\geq 0),   
      \ear
      \bear
      &&\sum \limits_{l=0}^{\infty} \frac{\left(l+1/2\right)^{2m+1}}
      {\left[u^2+\mu^2+\left(l+1/2\right)^2\right]^{(2n+1)/2}} \nn &&
      =\frac{(2m+1)!!}{(2n-1)!!}\left( -\frac{\partial }{\mu\partial \mu} 
   \right)^{n-m-1}\sum \limits_{l=0}^{\infty} \frac{\left(l+1/2\right)^{2m+1}}
      {\left[u^2+\mu^2+\left(l+1/2\right)^2\right]^{(2m+3)/2}}\nn
     &&=\frac{2 (-1)^{n-m}}{(2n-1)!!}\left( \frac{\partial }{\mu\partial \mu} 
      \right)\int \nolimits_{0}^{\sqrt{u^2+\mu^2}} 
      \frac{xdx}{\sqrt{u^2+\mu^2-x^2}}\left( \frac{d}{xdx} \right)^{m+1}
      \frac{x^{2m}}{1+e^{2\pi x}} \quad
      \left(m \geq 0, \atop n \geq m+2\right).  
      \ear

Now we can calculate the integrals over $u$ in (\ref{intsums}).
Note that these expressions must be expanded in powers of
$\varepsilon$:  
      \bear &&
      \int \nolimits_{0}^{\infty}\cos (\varepsilon u) \sqrt{u^2+\mu^2}du
      =\left( -\frac{d^2}{d\varepsilon^2}+\mu^2 \right)\int 
      \nolimits_{0}^{\infty}
      \frac{\cos (\varepsilon u) du}{\sqrt{u^2+\mu^2}}
      =\left( -\frac{d^2}{d\varepsilon^2}+\mu^2 \right)K_0(\varepsilon \mu) 
      \nn &=&-\frac{\mu}{\varepsilon}K_1(\varepsilon\mu)
      =\left(-\frac{\mu^2}{2}-\frac{\mu^4}{16}\varepsilon^2 \right)
      \left(C+\frac12\ln\frac{(\varepsilon \mu)^2}{4}\right)
      -\frac{1}{\varepsilon^2}+\frac{\mu^2}{4}+\frac{5}{64}\mu^4
      \varepsilon^2 \nn &&+O\left(\varepsilon^4 \ln \left|\varepsilon 
      \right| \right), 
      \ear
      \bear &&
      \int \nolimits_{0}^{\infty}\cos (\varepsilon u) (u^2+\mu^2)^{3/2}du
      =\left( -\frac{d^2}{d\varepsilon^2}+\mu^2 \right)\int 
      \nolimits_{0}^{\infty}
      \cos (\varepsilon u) \sqrt{u^2+\mu^2} du\nn
      &=&\frac{6}{\varepsilon^4}-\frac{3\mu^2}{2\varepsilon^2}-\frac38\mu^4 
      \left(C+\frac12\ln\frac{(\varepsilon \mu)^2}{4}\right)+\frac{9}{32}\mu^4
      +O\left(\varepsilon^2 \ln \left|\varepsilon \right| \right), 
      \ear
where $K_n(x)$ is Macdonald's function and $C$ is Euler's constant.

The second type of integral over $u$ has the form
      \bear
      I_{\pm} =\int \nolimits_{0}^{\infty}du \cos (\varepsilon u)
      \int \nolimits_{0}^{\sqrt{u^2+\mu^2}} 
      f(x) (u^2+\mu^2-x^2)^{\pm 1/2} \ dx .
      \ear
Changing the order of integration over $u$ and $x$ gives
      \bear
      I_-&=&\int \nolimits_{0}^{\mu}dx f(x)\int \nolimits_{0}^{\infty}du  
      \frac{\cos (\varepsilon u)}{\sqrt{u^2+\mu^2-x^2}}+
      \int \nolimits_{\mu}^{\infty}dx f(x) \int 
      \nolimits_{\sqrt{x^2-\mu^2}}^{\infty}
      du \frac{\cos (\varepsilon u)}{\sqrt{u^2+\mu^2-x^2}}.
      \ear
Since
      \bear \label{A}
      &&\int \nolimits_{0}^{\infty}\frac{\cos (\varepsilon u) du}
      {\sqrt{u^2+\mu^2-x^2}}=K_0(\varepsilon \sqrt{\mu^2-x^2})\nn
      &&=-\left(1+\frac{\varepsilon^2 (\mu^2-x^2)}{4} \right)
      \left(C+\frac12\ln\left|\frac{\varepsilon^2 (\mu^2-x^2)}{4}
      \right|\right)
      +\frac{\varepsilon^2 (\mu^2-x^2)}{4}
      +O\left(\varepsilon^4 \ln \left|\varepsilon \right| \right), 
      \ear
      \bear
      &&\int \nolimits_{\sqrt{x^2-\mu^2}}^{\infty}\frac{\cos (\varepsilon u)}
     {\sqrt{u^2+\mu^2-x^2}}=-\frac{\pi}{2}N_0(\varepsilon \sqrt{x^2-\mu^2})\nn
      &&=-\left( 1-\frac{\varepsilon^2 (x^2-\mu^2)}{4} \right)
     \left(C+\frac12\ln\left|\frac{\varepsilon^2(x^2- \mu^2)}{4}\right|\right)
      -\frac{\varepsilon^2 (x^2-\mu^2)}{4}
      +O\left(\varepsilon^4 \ln \left| \varepsilon \right| \right), 
      \ear
where  $N_0(x)$ is Neumann's function,  one can obtain
      \bear
      I_-&=&\int \nolimits_{0}^{\infty}dx f(x)\left\{\left(-1
      +\frac{\varepsilon^2 (x^2-\mu^2)}{4} \right)
      \left(C+\frac12\ln\left|\frac{\varepsilon^2(x^2- \mu^2)}{4}
      \right|\right)-\frac{\varepsilon^2 (x^2-\mu^2)}{4} \right. \nn && \left.
      +O\left(\varepsilon^4 \ln \left|\varepsilon \right| \right)  
       \right\},
      \ear
      \bear
      I_+&=&-\frac{\partial^2}{\partial \varepsilon^2} I_-  
      +\int \nolimits_{0}^{\infty}du \cos (\varepsilon u)
      \int \nolimits_{0}^{\sqrt{u^2+\mu^2}}\frac{f(x)(\mu^2-x^2)}
      {\sqrt{u^2+\mu^2-x^2}} \ dx \nn
      &=&-\frac{\partial^2}{\partial \varepsilon^2} I_-
      +\int \nolimits_{0}^{\infty}dx f(x)(\mu^2-x^2)\left\{ -
     \left(C+\frac12\ln\left|\frac{\varepsilon^2(x^2- \mu^2)}{4}\right|\right)
      +O\left(\varepsilon^2 \ln \left|\varepsilon \right| \right)\right\}
     \nn &=&\int \nolimits_{0}^{\infty}dx f(x)\left\{-\frac{1}{\varepsilon^2} 
      +\frac12(x^2-\mu^2)\left(C+\frac12\ln\left|\frac{\varepsilon^2(x^2
      - \mu^2)}{4}\right|\right)-\frac{(x^2-\mu^2)}{4} \right. \nn && \left.
      +O\left(\varepsilon^2 \ln \left|\varepsilon \right| \right)\right\}.
      \ear
If we also take in consideration 
      \beq
    \int \nolimits_{0}^{\infty}\frac{x dx}{1+e^{2\pi x}}=\frac{1}{48}, \quad  
      \int \nolimits_{0}^{\infty}\frac{x^3 dx}{1+e^{2\pi x}}=\frac{7}{1920},
      \eeq
      \beq
      \frac{\partial}{\mu \partial x}f(\mu x) =\frac{\partial}
      {x\partial \mu}f(\mu x),
      \eeq
the resulting expresions for $S^{m}_{n}(\varepsilon, \mu)$ can be
presented as Eqs. (\ref{start})-(\ref{finish}).

\section*{Appendix B}

\setcounter{equation}{0}
\renewcommand{\theequation}{B\arabic{equation}}

\bear\label{wkb2_tt}
\left< T^{t}_{t} \right>_{ren}&=&
\frac1{4 \pi^2}\left\{
\left({\frac {\xi}{8r^2}}
-{\frac {1}{64r^2}}\right)m^2
+{\frac {3{\xi}^{2}}{8r^4}}
-{\frac {11\xi}{96r^4}}
+{\frac {79}{7680r^4}} 
+\left[
\frac{m^4}2
+\left({\frac {2\xi}{r^2}}-\frac1{4r^2}\right)m^2 \right.\right. \nn && \left.
-{\frac {\xi}{2r^4}}  
+{\frac {2{\xi}^{2}}{r^4}}
+\frac{1}{32r^4}\right]{ I_1(\mu)} 
+\left[
-\frac{m^4}2+\left(\frac1{4r^2}-{\frac {2\xi}{r^2}}\right)m^2
+\left({\frac {\xi}{2r^4}}-{\frac {2{\xi}^{2}}{r^4}} 
\right.\right. \nn && \left.\left. -\frac1{32r^4}\right) 
\right]{ I_2(\mu)}
+\left[
-\frac{m^4}8
+\left(\frac{1}{12r^2}
-{\frac {\xi}{2r^2}}\right) m^2 
+\left({\frac {\xi}{6r^4}}-{\frac {1}{60r^4}}\right.\right. \nn &&\left.\left.
-{\frac {{\xi}^{2}}{2r^4}}\right)
\right]\ln\sqrt{\frac{\mu^2}{\mds^2r^2}}
+\left[
{\frac {\xi f'(r^2)'}{12r^2f}}
-{\frac {f'^{2}}{2304f^2{\mu}^{2}}}
-{\frac {f''}{1152f{\mu}^{2}}}
-{\frac {f''}{48f}}
-{\frac {7\xi f'^{2}}{24f^2}}        \right. \nn && \left.
+{\frac {f'^{2}}{16f^2}}
-{\frac {\xi f'^{2}}{576f^2{\mu}^{2}}}
-{\frac {f'(r^2)'}{48r^2f}}
+{\frac {\xi f''}{12f}}
-{\frac {\xi f''}{288f{\mu}^{2}}}\right] m^2    
+\left[
{\frac {f''}{4608fr^2{\mu}^{2}}}
-{\frac {(r^2)''}{48r^4}}                \right. \nn && \left.
-{\frac {(r^2)''}{4608r^4{\mu}^{2}}}
-{\frac {{\xi}^{2}f''}{144fr^2{\mu}^{2}}}  
-{\frac {\xi f''}{1152fr^2{\mu}^{2}}}
+{\frac {{\xi}^{2}f''}{8r^2f}}
-{\frac {\xi f''}{32r^2f}}
+{\frac {f''}{576r^2f}}
+{\frac {\xi(r^2)''}{4r^4}}                     \right. \nn && \left.
+{\frac {\xi(r^2)''}{1152r^4{\mu}^{2}}} 
+{\frac {{\xi}^{2}(r^2)''}{144r^4{\mu}^{2}}} 
-{\frac {3{\xi}^{2}(r^2)''}{4r^4}}
-{\frac {11\xi(r^2)'^{2}}{96r^6}}
-{\frac {5f'^{2}}{384r^2f^2}}
+{\frac {7f'{\xi}^{2}(r^2)'}{8r^4f}}                   \right. \nn && \left.
+{\frac {(r^2)'^{2}}{2304{\mu}^{2}r^6}}       
+{\frac {(r^2)'^{2}}{96r^6}}                  
-{\frac {29f'\xi(r^2)'}{96r^4f}}
+{\frac {5f'(r^2)'}{192r^4f}}     
+{\frac {11\xi f'^{2}}{64r^2f^2}}
-{\frac {9{\xi}^{2}f'^{2}}{16r^2f^2}}    
+{\frac {5{\xi}^{2}(r^2)'^{2}}{16r^6}}            \right. \nn && \left.
-{\frac {{\xi}^{2}(r^2)'^{2}}{72{\mu}^{2}r^6}}  
-{\frac {\xi(r^2)'^{2}}{576{\mu}^{2}r^6}}
+{\frac {f'^{2}}{9216f^2{\mu}^{2}r^2}}
-{\frac {5f'(r^2)'}{9216f{\mu}^{2}r^4}}
-{\frac {{\xi}^{2}f'^{2}}{288f^2{\mu}^{2}r^2}} 
-{\frac {\xi f'^{2}}{2304f^2{\mu}^{2}r^2}}        \right.  \nn && \left.
+{\frac {5f'{\xi}^{2}(r^2)'}{288f{\mu}^{2}r^4}} 
+{\frac {5f'\xi(r^2)'}{2304f{\mu}^{2}r^4}}\right]
+\left[
\left({\frac {\xi(r^2)''}{2r^2}}
+{\frac {(r^2)'^{2}}{48r^4}}
-{\frac {\xi(r^2)'^{2}}{8r^4}} 
-{\frac {(r^2)''}{12r^2}}\right) m^2    \right.  \nn && \left.
+{\frac {3{\xi}^{2}(r^2)'^{2}}{4r^6}}
-{\frac {\xi(r^2)'^{2}}{4r^6}}
+{\frac {(r^2)'^{2}}{48r^6}}
\right]{\ln\sqrt{\frac{\mu^2}{\mds^2r^2}}}
+\left[
\left({\frac {(r^2)'^{2}}{8r^4}}
-{\frac {3\xi(r^2)'^{2}}{4r^4}}\right) m^2            \right. \nn && \left.
-{\frac {3{\xi}^{2}(r^2)'^{2}}{2r^6}} 
-{\frac {(r^2)'^{2}}{32r^6}}
+{\frac {7\xi(r^2)'^{2}}{16r^6}}  
\right]{ I_1(\mu)} 
+\left[
\left(
{\frac {f'{\xi}^{2}(r^2)'}{r^2f}}
-{\frac {\xi f''}{2f}}
-{\frac {\xi(r^2)''}{r^2}}               \right. \right.  \nn && \left. \left.
+{\frac {f'^{2}}{32f^2}} 
+{\frac {2{\xi}^{2}(r^2)''}{r^2}}
+{\frac {f''}{16f}}
+{\frac {{\xi}^{2}f''}{f}}
-{\frac {f'(r^2)'}{16r^2f}}
+{\frac {3\xi(r^2)'^{2}}{8r^4}}
-{\frac {{\xi}^{2}f'^{2}}{2f^2}}
-{\frac {(r^2)'^{2}}{16r^4}}  \right.\right.  \nn && \left.\left.
+{\frac {(r^2)''}{8r^2}}
+{\frac {\xi f'^{2}}{4f^2}}
-{\frac {{\xi}^{2}(r^2)'^{2}}{2r^4}}\right) m^2
+{\frac {5{\xi}^{2}f'^{2}}{8r^2f^2}}
-{\frac {f''}{64r^2f}}
-{\frac {(r^2)''}{64r^4}}
-{\frac {{\xi}^{3}f'^{2}}{r^2f^2}}
+{\frac {7f'(r^2)'}{128r^4f}}       \right. \nn && \left.
-{\frac {3(r^2)'^{2}}{64r^6}}
+{\frac {4{\xi}^{3}(r^2)''}{r^4}}
+{\frac {2{\xi}^{3}f''}{r^2f}}
+{\frac {5\xi(r^2)''}{16r^4}}
-{\frac {5{\xi}^{2}f''}{4r^2f}}
-{\frac {17f'\xi(r^2)'}{32r^4f}}
+{\frac {\xi f''}{4r^2f}}               \right.  \nn && \left.
-{\frac {f'^{2}}{128r^2f^2}}
+{\frac {f'{\xi}^{2}(r^2)'}{2r^4f}}
-{\frac {9{\xi}^{2}(r^2)'^{2}}{8r^6}}  
+{\frac {2{\xi}^{3}f'(r^2)'}{r^4f}}
+{\frac {17\xi(r^2)'^{2}}{32r^6}}
-{\frac {{\xi}^{3}(r^2)'^{2}}{r^6}}  \right. \nn && \left.
-{\frac {2{\xi}^{2}(r^2)''}{r^4}} 
\right]{ \mu^2\left( \frac{\partial}{\mu\partial \mu}\right) I_0(\mu)}
+\left[
\left({\frac {\xi(r^2)''}{12r^2}}
-{\frac {\xi(r^2)'^{2}}{6r^4}}
-{\frac {f'\xi(r^2)'}{12r^2f}}
+{\frac {5\xi f'^{2}}{24f^2}} \right. \right.  \nn && \left.\left.
-{\frac {\xi f''}{6f}} 
+{\frac {f'(r^2)'}{48r^2f}}
+{\frac {f''}{24f}}
+{\frac {(r^2)'^{2}}{24r^4}}
-{\frac {(r^2)''}{48r^2}}
-{\frac {5f'^{2}}{96f^2}}\right) m^2  
-{\frac {(r^2)'^{2}}{384r^6}}
-{\frac {{\xi}^{2}f''}{3r^2f}} \right.  \nn && \left.
+{\frac {5f'^{2}}{384r^2f^2}} 
-{\frac {f'(r^2)'}{96r^4f}}
+{\frac {{\xi}^{2}(r^2)''}{3r^4}}
-{\frac {\xi(r^2)''}{8r^4}}
+{\frac {5{\xi}^{2}f'^{2}}{12r^2f^2}}
-{\frac {f''}{96r^2f}}
+{\frac {\xi f''}{8r^2f}}
+{\frac {(r^2)''}{96r^4}}                          \right.  \nn && \left. 
+{\frac {f'\xi(r^2)'}{8r^4f}}
-{\frac {{\xi}^{2}(r^2)'^{2}}{12r^6}}
+{\frac {\xi(r^2)'^{2}}{32r^6}}
-{\frac {5\xi f'^{2}}{32r^2f^2}}
-{\frac {f'{\xi}^{2}(r^2)'}{3r^4f}}  
\right]{ \mu^2\left( \frac{\partial}{\mu\partial \mu}\right)^2 \mu^2 I_0(\mu)}
\nn &&+\left[
\left({\frac {\xi f'^{2}}{16f^2}}
+{\frac {f'(r^2)'}{48r^2f}} 
-{\frac {(r^2)'^{2}}{192r^4}}
-{\frac {\xi f'(r^2)'}{12r^2f}}
+{\frac {\xi(r^2)'^{2}}{48r^4}}
-{\frac {f'^{2}}{64f^2}}\right) m^2
+\left({\frac {f'^{2}}{256r^2f^2}}    \right. \right.  \nn && \left. \left.
-{\frac {f'{\xi}^{2}(r^2)'}{4r^4f}} 
-{\frac {3\xi(r^2)'^{2}}{64r^6}}
+{\frac {{\xi}^{2}f'^{2}}{8r^2f^2}}
-{\frac {3\xi f'^{2}}{64r^2f^2}}      
+{\frac {(r^2)'^{2}}{256r^6}}
+{\frac {{\xi}^{2}(r^2)'^{2}}{8r^6}}
+{\frac {3\xi f'(r^2)'}{32r^4f}}   \right. \right.  \nn && \left. \left.
-{\frac {f'(r^2)'}{128r^4f}}\right)\right] 
{ \mu^2\left( \frac{\partial}{\mu\partial \mu}\right)^3 \mu^4 I_0(\mu)}
+\left[
\left({\frac {\xi(r^2)''}{3r^2}}
-{\frac {7(r^2)'^{2}}{96r^4}}
-{\frac {{\xi}^{2}(r^2)''}{r^2}} 
+{\frac {7f'(r^2)'}{96r^2f}}    \right. \right. \nn && \left. \left.
-{\frac {(r^2)''}{24r^2}}            
-{\frac {f'^{2}}{192f^2}}
+{\frac {\xi f''}{4f}}
+{\frac {{\xi}^{2}f'^{2}}{4f^2}}
+{\frac {{\xi}^{2}(r^2)'^{2}}{4r^4}}
-{\frac {{\xi}^{2}f''}{2f}}
-{\frac {\xi f'(r^2)'}{3r^2f}}
-{\frac {{\xi}^{2}f'(r^2)'}{2r^2f}} \right.\right. \nn && \left.
+{\frac {19\xi(r^2)'^{2}}{48r^4}}        
-{\frac {f''}{32f}}
-{\frac {\xi f'^{2}}{12f^2}}\right)m^2  
+{\frac {47f'\xi(r^2)'}{192r^4f}}
-{\frac {2{\xi}^{3}(r^2)''}{r^4}}
+{\frac {f'^{2}}{768r^2f^2}}
-{\frac {11{\xi}^{2}f'^{2}}{48r^2f^2}} \nn &&
+{\frac {f''}{128r^2f}} 
+{\frac {{\xi}^{3}f'^{2}}{2r^2f^2}}
-{\frac {f'{\xi}^{2}(r^2)'}{4r^4f}} 
-{\frac {5\xi(r^2)''}{32r^4}}
-{\frac {\xi f''}{8r^2f}}
+{\frac {5{\xi}^{2}f''}{8r^2f}}
+{\frac {{\xi}^{3}(r^2)'^{2}}{2r^6}}
-{\frac {{\xi}^{3}f''}{r^2f}}                   \nn &&
-{\frac {49\xi(r^2)'^{2}}{192r^6}}
-{\frac {{\xi}^{3}f'(r^2)'}{r^4f}}
+{\frac {(r^2)'^{2}}{48r^6}}
+{\frac {(r^2)''}{128r^4}}
+{\frac {{\xi}^{2}(r^2)''}{r^4}}   
+{\frac {31{\xi}^{2}(r^2)'^{2}}{49r^6}}
+\frac{\xi f'^2}{96r^2 f^2}  \nn && \left.
-{\frac {17f'(r^2)'}{768r^4f}} \right]
{2\left( \frac{\partial}{\mu\partial \mu}\right) \mu^2 I_1(\mu)}
+\left[
\left(
{\frac {\xi f'(r^2)'}{16r^2f}}
-{\frac {\xi f'^{2}}{16f^2}}
-{\frac {\xi(r^2)''}{24r^2}}
-{\frac {f'(r^2)'}{192r^2f}}
+{\frac {\xi f''}{24f}}                \right. \right. \nn && \left. \left.
+{\frac {f'^{2}}{96f^2}}         
-{\frac {(r^2)'^{2}}{192r^4}}
+{\frac {(r^2)''}{96r^2}}
-{\frac {f''}{96f}}\right) m^2
+{\frac {f''}{384r^2f}}
-{\frac {(r^2)''}{384r^4}}
+{\frac {f'(r^2)'}{768r^4f}}
-{\frac {f'^{2}}{384r^2f^2}}                  \right.  \nn && \left.
-{\frac {{\xi}^{2}f'^{2}}{8r^2f^2}}      
+{\frac {\xi(r^2)''}{32r^4}}
+{\frac {(r^2)'^{2}}{768r^6}} 
-{\frac {5\xi f'(r^2)'}{192r^4f}}
+{\frac {7\xi f'^{2}}{192r^2f^2}}
-{\frac {{\xi}^{2}(r^2)''}{12r^4}}
+{\frac {f'{\xi}^{2}(r^2)'}{8r^4f}}     \right.  \nn && \left.
-{\frac {\xi f''}{32r^2f}}
+{\frac {{\xi}^{2}f''}{12r^2f}} 
\right]{2\left( \frac{\partial}{\mu\partial \mu}\right)^2 \mu^4 I_1(\mu)}
+\left[
\left({\frac {(r^2)'^{2}}{384r^4}}
-{\frac {\xi f'^{2}}{96f^2}}
+{\frac {f'^{2}}{384f^2}}    \right. \right.  \nn && \left.\left.
-{\frac {f'(r^2)'}{192r^2f}}
-{\frac {\xi(r^2)'^{2}}{96r^4}}
+{\frac {\xi f'(r^2)'}{48r^2f}}\right) m^2  
+\left(
{\frac {f'(r^2)'}{768r^4f}}
-{\frac {(r^2)'^{2}}{1536r^6}}
+{\frac {\xi(r^2)'^{2}}{128r^6}} 
-{\frac {\xi f'(r^2)'}{64r^4f}}   \right. \right.   \nn && \left. \left.
+{\frac {\xi f'^{2}}{128r^2f^2}}
-{\frac {{\xi}^{2}f'^{2}}{48r^2f^2}}
-{\frac {{\xi}^{2}(r^2)'^{2}}{48r^6}}
-{\frac {f'^{2}}{1536r^2f^2}}
+{\frac {f'{\xi}^{2}(r^2)'}{24r^4f}}\right) \right]
{2\left( \frac{\partial}{\mu\partial \mu}\right)^3 \mu^6 I_1(\mu)} \nn &&
\left[ \left(
{\frac {r^2f'^{2}}{192f^2}}
-{\frac {f'(r^2)'}{192f}}
-{\frac {\xi f'(r^2)'}{48f}}
+{\frac {r^2\xi f'^{2}}{48f^2}}\right) m^4
+\left({\frac {{\xi}^{2}f'^{2}}{12f^2}}
-{\frac {(r^2)'^{2}}{768r^4}}
-{\frac {f'^{2}}{384f^2}}\right. \right. \nn && \left. \left.
+{\frac {\xi f'^{2}}{96f^2}} 
-{\frac {{\xi}^{2}f'(r^2)'}{8r^2f}}
+{\frac {\xi(r^2)'^{2}}{192r^4}}
-{\frac {\xi f'(r^2)'}{64r^2f}}
+{\frac {{\xi}^{2}(r^2)'^{2}}{24r^4}}
+{\frac {f'(r^2)'}{256r^2f}}\right) m^2
+\left(
{\frac {(r^2)'^{2}}{3072r^6}} \right. \right. \nn && \left. \left.
-{\frac {{\xi}^{3}f'(r^2)'}{6r^4f}}
-{\frac {\xi(r^2)'^{2}}{256r^6}}
+{\frac {{\xi}^{3}f'^{2}}{12r^2f^2}}
+{\frac {\xi f'(r^2)'}{128r^4f}}
+{\frac {{\xi}^{3}(r^2)'^{2}}{12r^6}}
-{\frac {f'(r^2)'}{1536r^4f}}
+{\frac {f'^{2}}{3072r^2f^2}} \right. \right. \nn && \left. \left.
-{\frac {\xi f'^{2}}{256r^2f^2}}\right) 
\right]{2\left( \frac{\partial}{\mu\partial \mu}\right)^2 \mu^2 I_1(\mu)}
+\left[ \left(
{\frac {r^2f''}{16f}}
-{\frac {5r^2f'^{2}}{64f^2}}
+{\frac {5r^2\xi f'^{2}}{16f^2}}
-{\frac {r^2\xi f''}{4f}} \right) m^4 \right. \nn && \left.
+\left(
{\frac {{\xi}^{2}(r^2)''}{2r^2}}
+{\frac {(r^2)''}{64r^2}}
+{\frac {3\xi f'(r^2)'}{16r^2f}}
-{\frac {(r^2)'^{2}}{32r^4}}
-{\frac {{\xi}^{2}f'(r^2)'}{2r^2f}}
-{\frac {3\xi(r^2)''}{16r^2}} 
+{\frac {5f'^{2}}{128f^2}} \right.  \right. \nn && \left. \left.
+{\frac {5{\xi}^{2}f'^{2}}{4f^2}}  
-{\frac {f''}{32f}}
-{\frac {15\xi f'^{2}}{32f^2}}
-{\frac {f'(r^2)'}{64r^2f}}
+{\frac {3\xi(r^2)'^{2}}{8r^4}}
-{\frac {{\xi}^{2}(r^2)'^{2}}{r^4}}
+{\frac {3\xi f''}{8f}}
-{\frac {{\xi}^{2}f''}{f}}\right) m^2 \right.  \nn && 
-{\frac {5\xi f'(r^2)'}{64r^4f}}
+{\frac {(r^2)'^{2}}{1024r^6}}
+{\frac {5\xi(r^2)''}{64r^4}}
-{\frac {5\xi f''}{64r^2f}}
-{\frac {(r^2)''}{256r^4}}
+{\frac {f''}{256r^2f}}
-{\frac {{\xi}^{3}(r^2)'^{2}}{4r^6}} \nn &&
-{\frac {5f'^{2}}{1024r^2f^2}} 
-{\frac {5\xi(r^2)'^{2}}{256r^6}} 
-{\frac {{\xi}^{3}f''}{r^2f}}
+{\frac {{\xi}^{2}f'(r^2)'}{2r^4f}}
+{\frac {{\xi}^{2}(r^2)'^{2}}{8r^6}}
-{\frac {{\xi}^{3}f'(r^2)'}{r^4f}}
+{\frac {{\xi}^{3}(r^2)''}{r^4}}  \nn && \left.
-{\frac {5{\xi}^{2}f'^{2}}{8r^2f^2}}
+{\frac {5{\xi}^{3}f'^{2}}{4r^2f^2}}
-{\frac {{\xi}^{2}(r^2)''}{2r^4}}
+{\frac {25\xi f'^{2}}{256r^2f^2}}
+{\frac {{\xi}^{2}f''}{2r^2f}}
+{\frac {f'(r^2)'}{256r^4f}}\right]
{S^0_2(\varepsilon, \mu)}_{\displaystyle|_{ \varepsilon=0}} \nn &&
+\left[
\left({\frac {15r^2\xi f'^{2}}{16f^2}}
+{\frac {5f'(r^2)'}{32f}}
-{\frac {5f'\xi(r^2)'}{8f}}
-{\frac {15r^2f'^{2}}{64f^2}} \right) m^4
+\left({\frac {15f'^{2}}{128f^2}}        
+{\frac {15{\xi}^{2}f'^{2}}{4f^2}} \right.  \right. \nn && \left.
+{\frac {5{\xi}^{2}(r^2)'^{2}}{4r^4}}   
-{\frac {45\xi f'^{2}}{32f^2}}
-{\frac {5f'{\xi}^{2}(r^2)'}{r^2f}}
-{\frac {5f'(r^2)'}{32r^2f}}
+{\frac {5(r^2)'^{2}}{128r^4}}
+{\frac {15f'\xi(r^2)'}{8r^2f}} \right. \nn && \left.
-{\frac {15\xi(r^2)'^{2}}{32r^4}}\right) m^2   
+{\frac {15f'{\xi}^{2}(r^2)'}{4r^4f}}
-{\frac {75f'\xi(r^2)'}{128r^4f}}
-{\frac {15{\xi}^{2}f'^{2}}{8r^2f^2}}
+{\frac {15f'(r^2)'}{512r^4f}}    
+{\frac {15{\xi}^{3}f'^{2}}{4r^2f^2}}  \nn && 
+{\frac {15{\xi}^{3}(r^2)'^{2}}{4r^6}}
-{\frac {15f'^{2}}{1024r^2f^2}}    
-{\frac {15{\xi}^{2}(r^2)'^{2}}{8r^6}}
+{\frac {75\xi(r^2)'^{2}}{256r^6}}
-{\frac {15{\xi}^{3}f'(r^2)'}{2r^4f}}
-{\frac {15(r^2)'^{2}}{1024r^6}}  \nn && \left.
+{\frac {75\xi f'^{2}}{256r^2f^2}}
\right]{ S^1_3(\varepsilon, \mu)}_{\displaystyle|_{ \varepsilon=0}} 
+\left[
\left ({\frac {5r^4\xi f'^{2}}{16f^2}} 
-{\frac {5r^4f'^{2}}{64f^2}}\right ){m}^{6}
+ \left(
{\frac {15\xi f'(r^2)'}{32f}} 
+{\frac {15\xi f'(r^2)'}{32f}}  \right.  \right. \nn && \left. \left.
-{\frac {5f'{\xi}^{2}(r^2)'}{4f}}
-{\frac {45r^2\xi f'^{2}}{64f^2}}
+{\frac {15r^2f'^{2}}{256f^2}}
-{\frac {5f'(r^2)'}{128f}}
+{\frac {15r^2{\xi}^{2}f'^{2}}{8f^2}}\right) m^4  
+\left(
{\frac {25\xi(r^2)'^{2}}{256r^4}} \right.  \right. \nn &&
-{\frac {15{\xi}^{2}f'^{2}}{8f^2}}
-{\frac {5{\xi}^{3}f'(r^2)'}{r^2f}}
-{\frac {25\xi f'(r^2)'}{64r^2f}}
-{\frac {15f'^{2}}{1024f^2}}
-{\frac {5{\xi}^{2}(r^2)'^{2}}{8r^4}} 
+{\frac {75\xi f'^{2}}{256f^2}}
+{\frac {5f'(r^2)'}{256r^2f}}        \nn && \left.
+{\frac {15{\xi}^{3}f'^{2}}{4f^2}}
+{\frac {5{\xi}^{2}f'(r^2)'}{2r^2f}}
+{\frac {5{\xi}^{3}(r^2)'^{2}}{4r^4}}
-{\frac {5(r^2)'^{2}}{1024r^4}}\right) m^2
+{\frac {5(r^2)'^{2}}{4096r^6}}
-{\frac {35\xi f'^{2}}{1024r^2f^2}}                         \nn &&
-{\frac {45f'{\xi}^{2}(r^2)'}{64r^4f}}
+{\frac {45{\xi}^{2}f'^{2}}{128r^2f^2}}
+{\frac {5{\xi}^{4}(r^2)'^{2}}{2r^6}}
-{\frac {25{\xi}^{3}f'^{2}}{16r^2f^2}}
+{\frac {35\xi f'(r^2)'}{512r^4f}}
+{\frac {25{\xi}^{3}f'(r^2)'}{8r^4f}}                      \nn && 
+{\frac {5f'^{2}}{4096r^2f^2}} 
+{\frac {45{\xi}^{2}(r^2)'^{2}}{128r^6}}
-{\frac {5{\xi}^{4}f'(r^2)'}{r^4f}}
-{\frac {5f'(r^2)'}{2048r^4f}}
+{\frac {5{\xi}^{4}f'^{2}}{2r^2f^2}}
-{\frac {35\xi(r^2)'^{2}}{1024r^6}} \nn && \left.
-{\frac {25{\xi}^{3}(r^2)'^{2}}{16r^6}} \right]
{ S^0_3(\varepsilon, \mu)}_{\displaystyle|_{ \varepsilon=0}} 
+\left[ \left(
-{\frac {5r^2 \xi f'^{2}}{16f^2}}
-{\frac {5r^2f'^{2}}{64f^2}} \right) m^4
+\left({\frac {5f'^{2}}{128f^2}}
-{5\frac {\xi f'^{2}}{32f^2}}          \right.  \right. \nn && \left. \left. 
+{\frac {5{\xi}^{2}f'(r^2)'}{4r^2f}}
-{\frac {5f'(r^2)'}{128r^2f}}
+{\frac {5f'\xi(r^2)'}{32r^2f}}
-{\frac {5{\xi}^{2}f'^{2}}{4f^2}}\right) m^2  
+{\frac {15\xi (r^2)'^{2}}{256r^6}}
-{\frac {5(r^2)'^{2}}{1024r^6}}                 \right. \nn && \left.
-{\frac {5{\xi}^{3} f'^{2}}{4r^2f^2}}
-{\frac {5{\xi}^{3}(r^2)'^{2}}{4r^6}}
+{\frac {15\xi f'^{2}}{256r^2f^2}}
-{\frac {15f'\xi (r^2)'}{128r^4f}}
-{\frac {5f'^{2}}{1024 r^2 f^2}} 
+{\frac {5{\xi}^{3}f'(r^2)'}{2r^4f}}             \right. \nn && \left. \left. 
+{\frac {5f'(r^2)'}{512 r^4 f}}  \right]
\frac{\partial^2 }{\partial \varepsilon^2}
S^0_3(\varepsilon,\mu)_{\displaystyle|_{ \varepsilon=0}}
\right\},
\ear

\bear
\left< T^{\rho}_{\rho} \right>_{ren}&=&\frac{1}{4 \pi^2}\left\{
\left ({\frac {\xi}{8r^2}}
-{\frac {1}{64r^2}} \right ){m}^{2}
+{\frac {79}{7680 r^4}}
-{\frac {11\xi}{96r^4}}
+{\frac {3{\xi}^{2}}{8r^4}}
+\left[
\frac{m^4}2
+\left ({\frac {2\xi}{r^2}}
-\frac1{4 r^2}\right ){m}^{2}   \right.  \right.  \nn && \left.
+{\frac {2{\xi}^{2}}{r^4}}            
-{\frac {\xi}{2r^4}}
+\frac1{32 r^4}
\right]{ I_1(\mu)} 
+\left[
-\frac{m^4}{2}
+\left (
\frac1{4 r^2}
-{\frac {2\xi}{r^2}}
\right ){m}^{2}
-\frac{1}{32 r^4}
+{\frac {\xi}{2r^4}}   \right. \nn && \left.
-{\frac {2{\xi}^{2}}{r^4}}\right]{ I_2(\mu)}  
+\left[
-\frac{m^4}{8}
+\left (
\frac1{12 r^2}
-{\frac {\xi}{2r^2}}\right ){m}^{2}
-{\frac {1}{60r^4}}
+{\frac {\xi}{6r^4}}
-{\frac {{\xi}^{2}}{2r^4}}\right] {\ln \sqrt{\frac{\mu^2}{\mds^2r^2}}} \nn &&
+\left[
{\frac {f''}{1152f {\mu}^{2}}}
-{\frac {f'(r^2)'}{48f r^2}}
-{\frac {f''}{48f }}
+{\frac {f'^2}{2304f^2{\mu}^{2}}}\right]{m}^{2}    
-{\frac {f'^2}{9216f^2r^2{\mu}^{2}}}
+{\frac {f'(r^2)'\xi}{6f r^4}}             \nn &&
-{\frac {f'(r^2)'{\xi}^{2}}{4f r^4}}
-{\frac {(r^2)'^2}{2304r^6{\mu}^{2}}}
+{\frac {(r^2)''}{4608r^4{\mu}^{2}}}
-{\frac {5f'(r^2)'\xi}{1152f r^4{\mu}^{2}}}
+{\frac {5f'(r^2)'}{9216f r^4{\mu}^{2}}}     
+{\frac {f'^2\xi}{1152f^2r^2{\mu}^{2}}}  \nn &&
+{\frac {f'' \xi}{576f r^2{\mu}^{2}}}
-{\frac {f'(r^2)'}{48f r^4}}
+{\frac {f'^2\xi}{192f^2r^2}}
-{\frac {f'' \xi}{32f r^2}}
+{\frac {(r^2)'^2\xi}{288r^6{\mu}^{2}}}
-{\frac {(r^2)''\xi}{576r^4{\mu}^{2}}}
-{\frac {f''}{4608f r^2{\mu}^{2}}}                                   \nn &&
-{\frac {(r^2)'^2{\xi}^{2}}{8r^6}}
-{\frac {(r^2)'^2\xi}{96r^6}}
-{\frac {f'^2}{1152f^2r^2}}
+{\frac {f''}{192f r^2}}
+{\frac {(r^2)''\xi}{16r^4}}
+{\frac {(r^2)'^2}{192r^6}}
-{\frac {(r^2)''}{96r^4}} \nn &&
+\left[
\left (  {\frac {\xi\,f'(r^2)'}{4f r^2}}
-{\frac {(r^2)'^2}{48r^4}} 
+{\frac {\xi\,(r^2)'^2}{8r^4}}
-{\frac {f'(r^2)'}{24f r^2}}\right ){m}^{2}
+{\frac {\xi\,(r^2)'^2}{4r^6}}
-{\frac {3(r^2)'^2{\xi}^{2}}{4r^6}}    \right.  \nn && \left.
-{\frac {(r^2)'^2}{48r^6}}
\right]{\ln \sqrt{\frac{\mu^2}{\mds^2r^2}}} 
+\left[\left ({\frac {3(r^2)'^2\xi}{4r^4}}
-{\frac {(r^2)'^2}{8r^4}}\right ){m}^{2}
+{\frac {3(r^2)'^2{\xi}^{2}}{2r^6}}
-{\frac {7(r^2)'^2\xi}{16r^6}}                         \right.  \nn && \left.
+{\frac {(r^2)'^2}{32r^6}}\right]{ I_1(\mu)}
+\left[
\left (
{\frac {(r^2)''}{8r^2}}
-{\frac {f''}{16f }}
-{\frac {\xi\,f''}{4f }}
-{\frac {\xi\,f'^2}{8f^2}}
-{\frac {f'^2}{32f^2}} 
-{\frac {3\xi\,f'(r^2)'}{4f r^2}}
-{\frac {\xi\,(r^2)''}{2r^2}}     \right. \right.    \nn && \left. \left.
+{\frac {\xi\,(r^2)'^2}{8r^4}}
+{\frac {3f'(r^2)'}{16f r^2}}
-{\frac {(r^2)'^2}{16r^4}}\right ){m}^{2}
+{\frac {5{\xi}^{2}(r^2)'^2}{4r^6}}
+{\frac {13f'(r^2)'\xi}{16f r^4}}
+{\frac {f'^2}{128f^2r^2}}
-{\frac {{\xi}^{2}f'^2}{4f^2r^2}}                 \right.  \nn && \left.
-{\frac {3(r^2)''}{64r^4}}
-{\frac {11f'(r^2)'}{128f r^4}}
-{\frac {\xi\,f'^2}{32f^2r^2}}
-{\frac {25(r^2)'^2\xi}{32r^6}}
-{\frac {{\xi}^{2}f''}{2f r^2}}
+{\frac {\xi\,(r^2)''}{2r^4}}
-{\frac {{\xi}^{2}(r^2)''}{r^4}}
+{\frac {5(r^2)'^2}{64r^6}}                           \right.  \nn && \left.
-{\frac {f'(r^2)'{\xi}^{2}}{f r^4}}
+{\frac {f''}{64f r^2}}
-{\frac {\xi\,f''}{16f r^2}}\right]  
{ \mu^2\left( \frac{\partial}{\mu\partial \mu}\right) I_0(\mu)} 
+\left[\left (
{\frac {(r^2)'^2}{24r^4}}
-{\frac {f'(r^2)'}{8f r^2}}
+{\frac {f''}{24f }}              \right. \right.  \nn && \left.\left.
+{\frac {3f'^2}{32f^2}}
-{\frac {(r^2)''}{48r^2}}\right ){m}^{2}
+{\frac {(r^2)''}{96r^4}}
+{\frac {3\xi\,f'^2}{16f^2r^2}}
-{\frac {f'(r^2)'\xi}{2f r^4}}
-{\frac {5(r^2)'^2}{128r^6}}
+{\frac {5(r^2)'^2\xi}{16r^6}}
-{\frac {f''}{96f r^2}}         \right.  \nn && \left.
-{\frac {3f'^2}{128f^2r^2}} 
+{\frac {f'(r^2)'}{16f r^4}}
-{\frac {\xi\,(r^2)''}{12r^4}} 
+{\frac {\xi\,f''}{12f r^2}}\right]
{ \mu^2\left( \frac{\partial}{\mu\partial \mu}\right)^2 \mu^2 I_0(\mu)}
+\left[ \left (
{\frac {f'(r^2)'}{48f r^2}}   \right.  \right.  \nn && \left.\left.
-{\frac {(r^2)'^2}{192r^4}}
-{\frac {f'^2}{64f^2}}\right ){m}^{2}
+{\frac {f'(r^2)'\xi}{16f r^4}}
- {\frac {f'(r^2)'}{128f r^4}}
-{\frac {(r^2)'^2\xi}{32r^6}}
-{\frac {\xi\,f'^2}{32f^2r^2}}
+{\frac {(r^2)'^2}{256r^6}}                          \right.  \nn && \left.
+{\frac {f'^2}{256f^2r^2}}\right]
{ \mu^2\left( \frac{\partial}{\mu\partial \mu}\right)^3 \mu^4 I_0(\mu)}
+\left[
\left ( {\frac {\xi\,f''}{8f }}
+{\frac {\xi\,(r^2)''}{4r^2}}
+{\frac {f'^2}{192f^2}}
+{\frac {\xi\,f'^2}{16f^2}}    \right. \right.  \nn && \left. \left.
+{\frac {13(r^2)'^2}{96r^4}}
+{\frac {f''}{32f }}
-{\frac {5\xi\,(r^2)'^2}{16r^4}}
+{\frac {\xi\,f'(r^2)'}{4f r^2}}
-{\frac {(r^2)''}{12r^2}}
-{\frac {13f'(r^2)'}{96f r^2}}\right ){m}^{2}
+{\frac {{\xi}^{2}f''}{4f r^2}}                   \right.  \nn && \left.
+{\frac {29f'(r^2)'}{768f r^4}}
+{\frac {{\xi}^{2}(r^2)''}{2r^4}}
+{\frac {\xi\,f''}{32f r^2}} 
+{\frac {3(r^2)''}{128r^4}}
-{\frac {7(r^2)'^2}{192r^6}}
-{\frac {35\xi\,f'(r^2)'}{96f r^4}}
+{\frac {{\xi}^{2}f'^2}{8f^2r^2}}                \right.  \nn && \left.
-{\frac {5(r^2)'^2{\xi}^{2}}{8r^6}}
-{\frac {\xi\,f'^2}{192f^2r^2}}
+{\frac {71\xi\,(r^2)'^2}{192r^6}}
+{\frac {f'(r^2)'{\xi}^{2}}{2f r^4}}
-{\frac {\xi\,(r^2)''}{4r^4}} 
-{\frac {f'^2}{256f^2r^2}}                \right.  \nn && \left.
-{\frac {f''}{768f r^2}}\right]
{2\left( \frac{\partial}{\mu\partial \mu}\right) \mu^2 I_1(\mu)} 
+\left[ 
\left (
{\frac {(r^2)''}{96r^2}}
-{\frac {f'^2}{48f^2}}
-{\frac {f''}{96f }}
-{\frac {7(r^2)'^2}{192r^4}} \right. \right.  \nn && \left.\left.
+{\frac {11f'(r^2)'}{192f r^2}}\right ){m}^{2}
+{\frac {11\xi\,f'(r^2)'}{96f r^4}}
+{\frac {\xi\,(r^2)''}{48r^4}}
+{\frac {f''}{384f r^2}}
+{\frac {7(r^2)'^2}{768r^6}}
-{\frac {11f'(r^2)'}{768f r^4}}           \right.  \nn && \left.
+{\frac {f'^2}{192f^2r^2}}
-{\frac {(r^2)''}{384r^4}} 
-{\frac {7(r^2)'^2\xi}{96r^6}}
-{\frac {\xi\,f''}{48f r^2}} 
-{\frac {\xi\,f'^2}{24f^2r^2}}\right]
{2\left( \frac{\partial}{\mu\partial \mu}\right)^2 \mu^4 I_1(\mu)}  \nn && 
+\left[
\left (
{\frac {f'^2}{384f^2}} 
-{\frac {f'(r^2)'}{192f r^2}}
+{\frac {(r^2)'^2}{384r^4}}\right ){m}^{2}
+{\frac {\xi\,f'^2}{192f^2r^2}}
-{\frac {\xi\,f'(r^2)'}{96f r^4}}
-{\frac {f'^2}{1536f^2r^2}}
+{\frac {f'(r^2)'}{768f r^4}} \right.  \nn && \left.
-{\frac {(r^2)'^2}{1536r^6}} 
+{\frac {\xi\,(r^2)'^2}{192r^6}}\right]
{2\left( \frac{\partial}{\mu\partial \mu}\right)^3 \mu^6 I_1(\mu)}
+\left[
\left ( {\frac {f'(r^2)'}{192f }}
-{\frac {r^2f'^2}{192f^2}}\right ){m}^{4} 
+\left (
{\frac {f'^2}{384f^2}}        \right. \right.  \nn && \left.\left.
-{\frac {f'(r^2)'}{256f r^2}}
-{\frac {\xi\,(r^2)'^2}{96r^4}}
+{\frac {(r^2)'^2}{768r^4}}
-{\frac {\xi\,f'^2}{48f^2}}
+{\frac {\xi\,f'(r^2)'}{32f r^2}}\right ){m}^{2}
+{\frac {\xi\,(r^2)'^2}{192r^6}}
+{\frac {f'(r^2)'}{1536f r^4}}            \right.  \nn && \left.
+{\frac {f'(r^2)'{\xi}^{2}}{24f r^4}}
+{\frac {\xi\,f'^2}{192f^2r^2}}
-{\frac {\xi\,f'(r^2)'}{96f r^4}}
-{\frac {f'^2}{3072 f^2 r^2}}
-{\frac {(r^2)'^2}{3072r^6}}
-{\frac {(r^2)'^2{\xi}^{2}}{48r^6}}     \right.  \nn && \left.
-{\frac {{\xi}^{2}f'^2}{48f^2r^2}}\right]
{2\left( \frac{\partial}{\mu\partial \mu}\right)^2 \mu^2 I_1(\mu)}  
+\left[
\left ( {\frac {r^2f''}{16f }} 
+{\frac {9r^2f'^2}{64f^2}}\right ){m}^{4}
+\left ( {\frac {(r^2)''}{64r^2}}
+{\frac {\xi\,f''}{4f }}               \right.  \right.  \nn && \left.\left.
+{\frac {3f'(r^2)' }{32f r^2}}
-{\frac {f''}{32f }}
+{\frac {9\xi\,f'^2}{16f^2}}
+{\frac {\xi\,(r^2)'^2}{4r^4}}
-{\frac {\xi\,(r^2)''}{8r^2}}
-{\frac {(r^2)'^2}{32r^4}}
-{\frac {9f'^2}{128f^2}}     \right. \right. \nn && \left.\left.
-{\frac {3\xi\,f'(r^2)'}{4f r^2}}\right ){m}^{2}
+{\frac {9{\xi}^{2}f'^2}{16f^2r^2}}
-{\frac {9\xi\,f'^2}{64f^2r^2}}
+{\frac {f''}{256f r^2}}
+{\frac {\xi\,(r^2)''}{16r^4}}
+{\frac {15(r^2)'^2}{1024r^6}}
+{\frac {{\xi}^{2}f''}{4f r^2}}     \right. \nn && \left.
-{\frac {3(r^2)'f'  }{128f r^4}}
-{\frac {(r^2)''}{256r^4}}
-{\frac {\xi\,f''}{16f r^2}}
+{\frac {9f'^2}{1024f^2r^2}}
+{\frac {3\xi\,f'(r^2)'}{8f r^4}}
-{\frac {3(r^2)'f'{\xi}^{2}}{2f r^4}}
-{\frac {15\xi\,(r^2)'^2}{64r^6}}                    \right. \nn && \left.
+{\frac {15(r^2)'^2{\xi}^{2}}{16r^6}}
-{\frac {{\xi}^{2}(r^2)''}{4r^4}}\right]
{S^0_2(\varepsilon, \mu)}_{\displaystyle|_{ \varepsilon=0}} 
+\left[
\left (
{\frac {5f'(r^2)'}{32f }}
-{\frac {15r^2f'^2}{64f^2}}\right ){m}^{4}
+\left (
{\frac {15f'^2}{128f^2}}                   \right. \right. \nn && \left.\left.
-{\frac {15\xi\,f'^2}{16f^2}}
+{\frac {5(r^2)'^2}{128r^4}}
-{\frac {5(r^2)'f'  }{32f r^2}}
-{\frac {5(r^2)'^2\xi}{16r^4}}
+{\frac {5(r^2)'f'\xi}{4f r^2}}\right ){m}^{2}
+{\frac {15(r^2)'f'  }{512f r^4}}                    \right. \nn && \left.
-{\frac {15f'^2}{1024f^2r^2}}
-{\frac {15(r^2)'^2{\xi}^{2}}{16r^6}}
-{\frac {15(r^2)'f'\xi}{32f r^4}}
+{\frac {15\xi\,f'^2}{64f^2r^2}}
+{\frac {15(r^2)'^2\xi}{64r^6}}
-{\frac {15(r^2)'^2}{1024r^6}}                    \right. \nn && \left.
+{\frac {15(r^2)'f'{\xi}^{2}}{8f r^4}}
-{\frac {15{\xi}^{2}f'^2}{16f^2r^2}}\right]
{S^1_3(\varepsilon, \mu)}_{\displaystyle|_{ \varepsilon=0}} 
+\left[
\left ( {\frac {5\xi\,f'(r^2)'}{16f }}
-{\frac {15r^2\xi\,f'^2}{32f^2}}
-{\frac {5f'(r^2)'}{128f }}            \right.  \right. \nn && \left. \left.
+{\frac {15r^2f'^2}{256f^2}}\right ){m}^{4}
+\left ( {\frac {5\xi\,(r^2)'^2}{64r^4}}
+{\frac {15\xi\,f'^2}{64f^2}}
+{\frac {5f'(r^2)'}{256f r^2}}
-{\frac {5(r^2)'^2}{1024r^4}}
-{\frac {15f'^2}{1024f^2}}
-{\frac {15{\xi}^{2}f'^2}{16f^2}}      \right.  \right. \nn && \left. \left.
-{\frac {5\xi\,f'(r^2)'}{16f r^2}}
+{\frac {5{\xi}^{2}f'(r^2)'}{4f r^2}}
-{\frac {5(r^2)'^2{\xi}^{2}}{16r^4}}\right ){m}^{2}
-{\frac {15f'(r^2)'{\xi}^{2}}{32f r^4}}
-{\frac {5{m}^{6}r^4f'^2}{64f^2}}
+{\frac {5(r^2)'^2}{4096r^6}}                    \right. \nn && \left. 
-{\frac {5f'(r^2)'}{2048f r^4}}
-{\frac {5{\xi}^{3}f'^2}{8f^2r^2}}
-{\frac {5{\xi}^{3}(r^2)'^2}{8r^6}}
+{\frac {15(r^2)'^2{\xi}^{2}}{64r^6}}
-{\frac {15(r^2)'^2\xi}{512r^6}}
+{\frac {5{\xi}^{3}f'(r^2)'}{4f r^4}}
+{\frac {15{\xi}^{2}f'^2}{64f^2r^2}}          \right. \nn && \left.
+{\frac {5f'^2}{4096f^2r^2}}
+{\frac {15\xi\,f'(r^2)'}{256f r^4}}
-{\frac {15\xi\,f'^2}{512f^2r^2}}\right]
{S^0_3(\varepsilon, \mu)}_{\displaystyle|_{ \varepsilon=0}}  
+\left[
\left (
{\frac {5\xi\,f'^2}{16f^2}}
+{\frac {5f'(r^2)'}{128f r^2}}         \right. \right. \nn && \left.\left. 
-{\frac {5f'^2}{128f^2}}
-{\frac {5f'(r^2)'\xi}{16f r^2}}\right ){m}^{2}
+{\frac {5f'^2}{1024f^2r^2}}
-{\frac {5(r^2)'^2\xi}{64r^6}}
+{\frac {5(r^2)'^2}{1024r^6}}
+{\frac {5{\xi}^{2}f'^2}{16f^2r^2}}\right. \nn && \left. \left.
+{\frac {5(r^2)'^2{\xi}^{2}}{16r^6}} 
-{\frac {5f'(r^2)'{\xi}^{2}}{8f r^4}} 
-{\frac {5\xi\,f'^2}{64f^2r^2}}
+{\frac {5f'(r^2)'\xi}{32f r^4}}
+{\frac {5r^2f'^2{m}^{4}}{64f^2}} \right. \right.  \nn && \left. \left. \left.
-{\frac {5f'(r^2)'}{512f r^4}}\right]
\frac{\partial^2 }{\partial \varepsilon^2}
S^0_3(\varepsilon,\mu)_{\displaystyle|_{ \varepsilon=0}} \right\}
\right\},
\ear

\bear\label{wkb2_thth}
\left< T^{\theta}_{\theta} \right>_{ren}
&=&\left< T^{\varphi}_{\varphi} \right>_{ren}=
\frac1{4 \pi^2}\left\{ 
\left (
{\frac {\xi}{8r^2}}
-{\frac {1}{64r^2}} \right ){m}^{2}
-{\frac {1}{512r^4}}
+{\frac {\xi}{32r^4}}
-{\frac {{\xi}^{2}}{8r^4}}
+\left[
\left (
\frac{1}{8r^2}
-{\frac {\xi}{r^2}} \right ){m}^{2}   \right. \right. \nn && \left.
-{\frac {2{\xi}^{2}}{r^4}}
+{\frac {\xi}{2r^4}}  
-\frac{1}{32r^4}              \right]{I_1(\mu)} 
+\left[
\frac{m^4}2
+\left (
{\frac {2\xi}{r^2}}
-\frac{1}{4r^2}         \right ){m}^{2}
+{\frac {2{\xi}^{2}}{r^4}}
-{\frac {\xi}{2r^4}}
+\frac{1}{32r^4}\right]{I_2(\mu)} \nn &&
+\left[
-\frac{m^4}{8}
+{\frac {{\xi}^{2}}{2r^4}}
-{\frac {\xi}{6r^4}}                  
+{\frac {1}{60r^4}}\right]
{\ln \sqrt{\frac{\mu^2}{\mds^2r^2}}} 
+\left[
{\frac {f'^{2}}{16f^2}}
-{\frac {f'(r^2)'}{48r^2f}}
-{\frac {f''}{48f}}
+{\frac {f''}{1152f{\mu}^{2}}}                         \right. \nn && \left.
-{\frac {7\xi\,f'^{2}}{24f^2}}
+{\frac {f'^{2}}{2304f^2{\mu}^{2}}}
+{\frac {\xi\,f''}{12f}}
-{\frac {\xi\,f''}{288f{\mu}^{2}}}
+{\frac {\xi\,f'(r^2)'}{12r^2f}}
-{\frac {\xi\,f'^{2}}{576f^2{\mu}^{2}}}                    \right]{m}^{2}
-{\frac {f''}{4608fr^2{\mu}^{2}}}                                   \nn && 
-{\frac {\xi\,(r^2)''}{384r^4{\mu}^{2}}}
+{\frac {5f'(r^2)'}{9216f{\mu}^{2}r^4}}
+{\frac {\xi\,(r^2)'^2}{192{\mu}^{2}r^6}}
-{\frac {f'^{2}}{9216f^2{\mu}^{2}r^2}} 
-{\frac {{\xi}^{2}f'^{2}}{288f^2{\mu}^{2}r^2}}
+{\frac {\xi\,f'^{2}}{768f^2{\mu}^{2}r^2}}                        \nn && 
-{\frac {{\xi}^{2}(r^2)'^2}{72{\mu}^{2}r^6}}
-{\frac {15\xi\,f'(r^2)'}{64r^4f}}
-{\frac {11\xi\,(r^2)'^2}{96r^6}}
+{\frac {5{\xi}^{2}(r^2)'^2}{16r^6}}
+{\frac {3f'{\xi}^{2}(r^2)'}{4r^4f}}
-{\frac {1}{96}}\,{\frac {f''}{r^2f}}
+{\frac {5\xi\,f'^{2}}{48r^2f^2}}                                    \nn &&
+{\frac {7(r^2)'f'}{384r^4f}}
-{\frac {{\xi}^{2}(r^2)''}{2r^4}}
+{\frac {17\xi\,(r^2)''}{96r^4}}
-{\frac {{\xi}^{2}f''}{8r^2f}}
+{\frac {\xi\,f''}{12r^2f}}
-{\frac {f'^{2}}{192r^2f^2}}
-{\frac {7{\xi}^{2}f'^{2}}{16r^2f^2}} 
-{\frac {(r^2)''}{64r^4}}                        \nn &&
+{\frac {(r^2)'^2}{96r^6}}
+{\frac {(r^2)''}{4608r^4{\mu}^{2}}}
-{\frac {5f'(r^2)'\xi}{768f{\mu}^{2}r^4}}
+{\frac {5{\xi}^{2}f'(r^2)'}{288f{\mu}^{2}r^4}}
-{\frac {(r^2)'^2}{2304{\mu}^{2}r^6}}
-{\frac {{\xi}^{2}f''}{144fr^2{\mu}^{2}}}                       \nn &&
+{\frac {\xi\,f''}{384fr^2{\mu}^{2}}}
+{\frac {{\xi}^{2}(r^2)''}{144r^4{\mu}^{2}}} 
+\left[
\left (
{\frac {\xi\,(r^2)''}{4r^2}}
-{\frac {\xi\,(r^2)'^2}{8r^4}}
+{\frac {f'^{2}}{48f^2}}
-{\frac {\xi\,f'^{2}}{8f^2}}
+{\frac {(r^2)'^2}{48r^4}}
-{\frac {(r^2)''}{24r^2}}   \right. \right.  \nn && \left. \left.
+{\frac {\xi\,f'(r^2)'}{8r^2f}}
+{\frac {\xi\,f''}{4f}}
-{\frac {f''}{24f}}
-{\frac {(r^2)'f'}{48r^2f}}                            \right ){m}^{2} 
+{\frac {3{\xi}^{2}(r^2)'^2}{2r^6}}
-{\frac {(r^2)'f'}{48r^4f}}
-{\frac {3f'{\xi}^{2}(r^2)'}{4r^4f}}              \right.  \nn && \left.
-{\frac {\xi\,(r^2)'^2}{2r^6}} 
-{\frac {(r^2)''}{24r^4}}
+{\frac {(r^2)'^2}{24r^6}}
+{\frac {\xi\,f'(r^2)'}{4r^4f}}
-{\frac {3{\xi}^{2}(r^2)''}{2r^4}}
+{\frac {\xi\,(r^2)''}{2r^4}}
\right]{\ln \sqrt{\frac{\mu^2}{\mds^2r^2}}} \nn &&
+\left[
\left (
{\frac {3\xi\,(r^2)'f'}{4r^2f}}
-{\frac {3\xi\,(r^2)'^2}{2r^4}}
+{\frac {3\xi\,(r^2)''}{2r^2}}
-{\frac {(r^2)'f'}{8r^2f}}
-{\frac {(r^2)''}{4r^2}}
+{\frac {(r^2)'^2}{4r^4}}                        \right ){m}^{2}
-{\frac {(r^2)'^2}{16r^6}}                       \right.  \nn && 
+{\frac {3f'{\xi}^{2}(r^2)'}{2r^4f}}
+{\frac {7\xi\,(r^2)'^2}{8r^6}}
+{\frac {(r^2)''}{16r^4}}
-{\frac {7\xi\,(r^2)''}{8r^4}}
+{\frac {(r^2)'f'}{32r^4f}}
-{\frac {3{\xi}^{2}(r^2)'^2}{r^6}}
+{\frac {3{\xi}^{2}(r^2)''}{r^4}}                     \nn && \left.
-{\frac {7\xi\,(r^2)'f'}{16r^4f}}       \right]{I_1(\mu)} 
+\left[
\left (
{\frac {2{\xi}^{2}(r^2)''}{r^2}}
+{\frac {3\xi\,(r^2)'^2}{8r^4}}
-{\frac {{\xi}^{2}(r^2)'^2}{2r^4}}
-{\frac {f'(r^2)'}{16r^2f}}
+{\frac {f'^{2}}{32f^2}}
+{\frac {{\xi}^{2}f''}{f}}  \right. \right. \nn && \left.
+{\frac {(r^2)''}{8r^2}}
-{\frac {\xi\,f''}{2f}} 
+{\frac {{\xi}^{2}f'(r^2)'}{r^2f}}
-{\frac {\xi\,(r^2)''}{r^2}}
-{\frac {{\xi}^{2}f'^{2}}{2f^2}}
+{\frac {f''}{16f}}
-{\frac {(r^2)'^2}{16r^4}}
+{\frac {\xi\,f'(r^2)'}{4r^2f}}     \right ){m}^{2} \nn && \left.
+{\frac {3\xi\,(r^2)'^2}{4r^6}}
-{\frac {27\xi\,f'(r^2)'}{32r^4f}}
-{\frac {{\xi}^{3}(r^2)'^2}{r^6}}
-{\frac {{\xi}^{2}f'^{2}}{8r^2f^2}}
-{\frac {f'^{2}}{128r^2f^2}}
+{\frac {4{\xi}^{3}(r^2)''}{r^4}}
-{\frac {f''}{64r^2f}}                                \right. \nn && \left.
+{\frac {3\xi\,f'^{2}}{32r^2f^2}}
+{\frac {2{\xi}^{3}f'(r^2)'}{r^4f}}
+{\frac {9(r^2)'f'}{128r^4f}}
-{\frac {3{\xi}^{2}f''}{4r^2f}}
+{\frac {3\xi\,f''}{16r^2f}}
-{\frac {\xi\,(r^2)''}{16r^4}}
-{\frac {{\xi}^{2}(r^2)''}{r^4}}
-{\frac {(r^2)'^2}{16r^6}}                    \right. \nn && \left.
+{\frac {(r^2)''}{64r^4}} 
+{\frac {2{\xi}^{3}f''}{r^2f}}
-{\frac {{\xi}^{3}f'^{2}}{r^2f^2}}
-{\frac {15{\xi}^{2}(r^2)'^2}{8r^6}}
+{\frac {2f'{\xi}^{2}(r^2)'}{r^4f}}
\right]{\mu^2\left( \frac{\partial}{\mu\partial \mu}\right) I_0(\mu)} 
+\left[
\left (
-{\frac {\xi\,f''}{6f}}                   \right. \right. \nn && \left.
-{\frac {f'^{2}}{16f^2}}
-{\frac {(r^2)''}{48r^2}}
+{\frac {\xi\,(r^2)''}{12r^2}}
+{\frac {(r^2)'^2}{24r^4}}
+{\frac {f''}{48f}}
-{\frac {\xi\,(r^2)'f'}{12r^2f}}
+{\frac {5\xi\,f'^{2}}{24f^2}}
-{\frac {\xi\,(r^2)'^2}{6r^4}}                   \right. \nn && \left.\left.
+{\frac {f'(r^2)'}{48r^2f}}                                 \right ){m}^{2}
-{\frac {13\xi\,(r^2)'^2}{96r^6}}
-{\frac {\xi\,(r^2)''}{24r^4}}
+{\frac {5{\xi}^{2}f'^{2}}{12r^2f^2}}
-{\frac {f'{\xi}^{2}(r^2)'}{3r^4f}}
-{\frac {7f'(r^2)'}{192r^4f}}                  \right. \nn && \left.
+{\frac {7(r^2)'^2}{384r^6}}
+{\frac {\xi\,f'(r^2)'}{3r^4f}}
-{\frac {{\xi}^{2}f''}{3r^2f}}
+{\frac {\xi\,f''}{24r^2f}} 
-{\frac {19\xi\,f'^{2}}{96r^2f^2}}
-{\frac {{\xi}^{2}(r^2)'^2}{12r^6}}
+{\frac {7f'^{2}}{384r^2f^2}}                       \right. \nn && \left.
+{\frac {{\xi}^{2}(r^2)''}{3r^4}}
\right]{ \mu^2\left( \frac{\partial}{\mu\partial \mu}\right)^2 \mu^2 I_0(\mu)}
+\left[
\left (
{\frac {\xi\,f'^{2}}{16f^2}}
+{\frac {f'(r^2)'}{96r^2f}}
-{\frac {f'^{2}}{192f^2}} 
-{\frac {(r^2)'^2}{192r^4}}
-{\frac {\xi\,f'(r^2)'}{12r^2f}}         \right. \right. \nn && \left. \left.
+{\frac {\xi\,(r^2)'^2}{48r^4}}                            \right ){m}^{2}
-{\frac {\xi\,f'^{2}}{64r^2f^2}}
+{\frac {{\xi}^{2}(r^2)'^2}{8r^6}}
+{\frac {{\xi}^{2}f'^{2}}{8r^2f^2}}
-{\frac {\xi\,(r^2)'^2}{64r^6}}
-{\frac {f'{\xi}^{2}(r^2)'}{4r^4f}}                   \right. \nn && \left.
+{\frac {\xi\,(r^2)'f'}{32r^4f}}
\right]{ \mu^2\left( \frac{\partial}{\mu\partial \mu}\right)^3 \mu^4 I_0(\mu)}
+\left[
\left (
{\frac {17f'(r^2)'}{96r^2f}}
+{\frac {7\xi\,(r^2)'^2}{12r^4}}
+{\frac {{\xi}^{2}(r^2)'^2}{4r^4}}
-{\frac {7(r^2)'^2}{48r^4}}                      \right. \right. \nn &&
+{\frac {\xi\,(r^2)''}{12r^2}}
-{\frac {f''}{32f}}
+{\frac {(r^2)''}{24r^2}} 
+{\frac {\xi\,f''}{8f}}
-{\frac {{\xi}^{2}f''}{2f}}
-{\frac {{\xi}^{2}f'(r^2)'}{2r^2f}}
-{\frac {{\xi}^{2}(r^2)''}{r^2}}
+{\frac {{\xi}^{2}f'^{2}}{4f^2}}
+{\frac {5\xi\,f'^{2}}{48f^2}}   \nn && \left.
-{\frac {17\xi\,f'(r^2)'}{24r^2f}}
-{\frac {5f'^{2}}{192f^2}}                      \right ){m}^{2}
+{\frac {13(r^2)'^2}{384r^6}} 
-{\frac {(r^2)''}{128r^4}}
+{\frac {7{\xi}^{2}f'^{2}}{48r^2f^2}}
-{\frac {5\xi\,f'^{2}}{64r^2f^2}}
+{\frac {5f'^{2}}{768r^2f^2}}                            \nn &&  
+{\frac {{\xi}^{3}(r^2)'^2}{2r^6}}
-{\frac {13\xi\,(r^2)'^2}{32r^6}}
+{\frac {{\xi}^{3}f'^{2}}{2r^2f^2}}
-{\frac {31f'(r^2)'}{768r^4f}}
+{\frac {49{\xi}^{2}(r^2)'^2}{48r^6}}
+{\frac {31\xi\,(r^2)'f'}{64r^4f}}
-{\frac {{\xi}^{3}f''}{r^2f}}                      \nn && \left. 
-{\frac {7f'{\xi}^{2}(r^2)'}{6r^4f}}
-{\frac {{\xi}^{3}f'(r^2)'}{r^4f}}
+{\frac {3{\xi}^{2}f''}{8r^2f}}
-{\frac {3\xi\,f''}{32r^2f}}
+{\frac {\xi\,(r^2)''}{32r^4}}
+{\frac {{\xi}^{2}(r^2)''}{2r^4}}
-{\frac {2{\xi}^{3}(r^2)''}{r^4}}                  \right. \nn && \left.
+{\frac {f''}{128r^2f}}
 \right]{2\left( \frac{\partial}{\mu\partial \mu}\right) \mu^2 I_1(\mu)} 
+\left[
\left (
{\frac {f'^{2}}{48f^2}}
+{\frac {\xi\,f'(r^2)'}{16r^2f}}
+{\frac {(r^2)'^2}{48r^4}}
-{\frac {\xi\,f'^{2}}{16f^2}}
-{\frac {f'(r^2)'}{24r^2f}}              \right. \right. \nn && \left.
-{\frac {\xi\,(r^2)''}{24r^2}}
+{\frac {\xi\,f''}{24f}}                             \right ){m}^{2} 
-{\frac {{\xi}^{2}f'^{2}}{8r^2f^2}}
-{\frac {f'^{2}}{192r^2f^2}}
-{\frac {(r^2)'^2}{192r^6}}
-{\frac {19\xi\,f'(r^2)'}{192r^4f}}
+{\frac {11\xi\,f'^{2}}{192r^2f^2}}                           \nn && 
-{\frac {{\xi}^{2}(r^2)''}{12r^4}} 
+{\frac {{\xi}^{2}f'(r^2)'}{8r^4f}}
+{\frac {f'(r^2)'}{96r^4f}}
+{\frac {{\xi}^{2}f''}{12r^2f}}  
+{\frac {\xi\,(r^2)'^2}{24r^6}}
-{\frac {\xi\,f''}{96r^2f}}                 \nn && \left.
+{\frac {\xi\,(r^2)''}{96r^4}}
 \right]{2\left( \frac{\partial}{\mu\partial \mu}\right)^2 \mu^4 I_1(\mu)}
+\left[
\left (
{\frac {\xi\,(r^2)'f'}{48fr^2}}
-{\frac {\xi\,(r^2)'^2}{96r^4}}
-{\frac {\xi\,f'^{2}}{96f^2}}                             \right ){m}^{2}
-{\frac {\xi\,(r^2)'f'}{192r^4f}}                 \right. \nn && \left.
+{\frac {\xi\,(r^2)'^2}{384r^6}}
+{\frac {\xi\,f'^{2}}{384r^2f^2}}
-{\frac {{\xi}^{2}(r^2)'^2}{48r^6}}
-{\frac {{\xi}^{2}f'^{2}}{48r^2f^2}}
+{\frac {f'{\xi}^{2}(r^2)'}{24r^4f}}
  \right]{2\left( \frac{\partial}{\mu\partial \mu}\right)^3 \mu^6 I_1(\mu)} 
\nn && +\left[
\left (
{\frac {f'(r^2)'}{192f}} 
-{\frac {\xi\,f'(r^2)'}{48f}}
+{\frac {\xi\,r^2f'^{2}}{48f^2}}
-{\frac {r^2f'^{2}}{192f^2}}                                \right ){m}^{4}
+\left (
{\frac {{\xi}^{2}f'^{2}}{12f^2}}
-{\frac {\xi\,(r^2)'^2}{64r^4}}
-{\frac {{\xi}^{2}f'(r^2)'}{8r^2f}}            \right. \right. \nn && \left.
-{\frac {f'(r^2)'}{256r^2f}}
+{\frac {3\xi\,f'(r^2)'}{64r^2f}} 
+{\frac {f'^{2}}{384f^2}}
+{\frac {(r^2)'^2}{768r^4}}
+{\frac {{\xi}^{2}(r^2)'^2}{24r^4}}
-{\frac {\xi\,f'^{2}}{32f^2}}                              \right ){m}^{2}
-{\frac {(r^2)'^2}{3072r^6}}                         \nn &&
-{\frac {{\xi}^{2}f'^{2}}{24r^2f^2}}
+{\frac {5\xi\,f'^{2}}{768r^2f^2}}
-{\frac {f'^{2}}{3072r^2f^2}}
+{\frac {{\xi}^{3}(r^2)'^2}{12r^6}}
+{\frac {5\xi\,(r^2)'^2}{768r^6}}
+{\frac {{\xi}^{3}f'^{2}}{12r^2f^2}}
+{\frac {f'(r^2)'}{1536r^4f}}                         \nn && \left. 
-{\frac {{\xi}^{2}(r^2)'^2}{24r^6}}
-{\frac {5\xi\,(r^2)'f'}{384r^4f}}
+{\frac {f'{\xi}^{2}(r^2)'}{12r^4f}}
-{\frac {{\xi}^{3}f'(r^2)'}{6r^4f}}
\right]{2\left( \frac{\partial}{\mu\partial \mu}\right)^2 \mu^2 I_1(\mu)}
+\left[
\left (
{\frac {r^2f''}{16f}}              \right. \right. \nn && \left. \left.
-{\frac {\xi\,r^2f''}{4f}}
-{\frac {5r^2f'^{2}}{64f^2}}
+{\frac {5\xi\,r^2f'^{2}}{16f^2}}\right ){m}^{4}
+\left (
{\frac {\xi\,f''}{4f}}
-{\frac {(r^2)'^2}{32r^4}}
-{\frac {f''}{64f}}
-{\frac {{\xi}^{2}f''}{f}}
+{\frac {(r^2)''}{64r^2}}        \right. \right. \nn &&
+{\frac {3f'^{2}}{64f^2}}
+{\frac {{\xi}^{2}(r^2)''}{2r^2}}
-{\frac {f'(r^2)'}{64r^2f}}
+{\frac {3\xi\,(r^2)'^2}{8r^4}}
+{\frac {5{\xi}^{2}f'^{2}}{4f^2}}
-{\frac {{\xi}^{2}(r^2)'^2}{r^4}}
-{\frac {17\xi\,f'^{2}}{32f^2}}               \nn && \left. 
+{\frac {3\xi\,f'(r^2)'}{16r^2f}}
-{\frac {{\xi}^{2}f'(r^2)'}{2r^2f}} 
-{\frac {3\xi\,(r^2)''}{16r^2}}                     \right ){m}^{2}
-{\frac {3{\xi}^{2}f'^{2}}{4r^2f^2}}
-{\frac {{\xi}^{3}(r^2)'^2}{4r^6}}
-{\frac {7f'^{2}}{1024r^2f^2}}                      \nn && 
-{\frac {3{\xi}^{2}(r^2)'^2}{8r^6}}
+{\frac {33\xi\,f'^{2}}{256r^2f^2}}
-{\frac {\xi\,f''}{64r^2f}}
-{\frac {{\xi}^{3}f'(r^2)'}{r^4f}}
+{\frac {\xi\,(r^2)''}{64r^4}}
-{\frac {15\xi\,(r^2)'f'}{64r^4f}}
-{\frac {{\xi}^{3}f''}{r^2f}}                      \nn && 
+{\frac {9f'{\xi}^{2}(r^2)'}{8r^4f}}
-{\frac {{\xi}^{2}(r^2)''}{4r^4}}
+{\frac {{\xi}^{3}(r^2)''}{r^4}}
+{\frac {27\xi\,(r^2)'^2}{256r^6}}
+{\frac {5{\xi}^{3}f'^{2}}{4r^2f^2}}
+{\frac {7(r^2)'f'}{512r^4f}}
-{\frac {7(r^2)'^2}{1024r^6}}                     \nn && \left.
+{\frac {{\xi}^{2}f''}{4r^2f}}
\right]{S^0_2(\varepsilon, \mu)}_{\displaystyle|_{ \varepsilon=0} } 
+\left[
\left (
{\frac {5(r^2)'f'}{32f}}
-{\frac {5r^2f'^{2}}{32f^2}}
-{\frac {5\xi\,f'(r^2)'}{8f}}
+{\frac {15\xi\,r^2f'^{2}}{16f^2}}\right ){m}^{4}\right. \nn && 
+\left (
{\frac {15{\xi}^{2}f'^{2}}{4f^2}} 
-{\frac {5(r^2)'f'}{64r^2f}}
-{\frac {25\xi\,f'^{2}}{32f^2}}
-{\frac {5f'{\xi}^{2}(r^2)'}{r^2f}}
+{\frac {5{\xi}^{2}(r^2)'^2}{4r^4}}
+{\frac {5f'^{2}}{128f^2}}
+{\frac {5(r^2)'^2}{128r^4}}                    \right. \nn && \left.
+{\frac {5\xi\,(r^2)'f'}{4r^2f}}
-{\frac {15\xi\,(r^2)'^2}{32r^4}}                            \right ){m}^{2}
+{\frac {15{\xi}^{2}f'(r^2)'}{8r^4f}}
+{\frac {15\xi\,(r^2)'^2}{256r^6}}
-{\frac {15{\xi}^{3}f'(r^2)'}{2r^4f}}
+{\frac {15{\xi}^{3}f'^{2}}{4r^2f^2}}                 \nn && \left.
-{\frac {15{\xi}^{2}(r^2)'^2}{16r^6}}
-{\frac {15\xi\,f'(r^2)'}{128r^4f}}
+{\frac {15{\xi}^{3}(r^2)'^2}{4r^6}} 
-{\frac {15{\xi}^{2}f'^{2}}{16r^2f^2}}
+{\frac {15\xi\,f'^{2}}{256r^2f^2}}
\right]{S^1_3(\varepsilon, \mu)}_{\displaystyle|_{ \varepsilon=0}} \nn &&
+\left[
\left (
{\frac {5\xi\,r^4f'^{2}}{16f^2}}
-{\frac {5r^4f'^{2}}{64f^2}}               \right ){m}^{6}
+\left (
{\frac {15\xi\,(r^2)'f'}{32f}}
-{\frac {5(r^2)'f'}{128f}}
-{\frac {5f'{\xi}^{2}(r^2)'}{4f}}
-{\frac {35\xi\,r^2f'^{2}}{64f^2}}  \right. \right. \nn &&  \left.
+{\frac {5r^2f'^{2}}{128f^2}}
+{\frac {15{\xi}^{2}r^2f'^{2}}{8f^2}}               \right ){m}^{4}
+\left (
{\frac {5f'(r^2)'}{512r^2f}}
-{\frac {5{\xi}^{2}(r^2)'^2}{8r^4}}
-{\frac {5{\xi}^{2}f'^{2}}{4f^2}}
-{\frac {5(r^2)'^2}{1024r^4}}                         \right. \nn &&
-{\frac {15\xi\,f'(r^2)'}{64r^2f}}
-{\frac {5{\xi}^{3}f'(r^2)'}{r^2f}}
+{\frac {15{\xi}^{2}f'(r^2)'}{8r^2f}}
+{\frac {25\xi\,(r^2)'^2}{256r^4}}
+{\frac {15{\xi}^{3}f'^{2}}{4f^2}}
-{\frac {5f'^{2}}{1024f^2}}                 \nn && \left.
+{\frac {35\xi\,f'^{2}}{256f^2}}
+{\frac {5{\xi}^{3}(r^2)'^2}{4r^4}}                       \right ){m}^{2}
+{\frac {5\xi\,(r^2)'f'}{512r^4f}}
-{\frac {15f'{\xi}^{2}(r^2)'}{64r^4f}}
-{\frac {15{\xi}^{3}f'^{2}}{16r^2f^2}}
-{\frac {5{\xi}^{4}f'(r^2)'}{r^4f}}                \nn &&
+{\frac {5{\xi}^{4}f'^{2}}{2r^2f^2}}
+{\frac {15{\xi}^{3}f'(r^2)'}{8r^4f}}
-{\frac {15{\xi}^{3}(r^2)'^2}{16r^6}}
-{\frac {5\xi\,(r^2)'^2}{1024r^6}} 
+{\frac {15{\xi}^{2}(r^2)'^2}{128r^6}}
+{\frac {5{\xi}^{4}(r^2)'^2}{2r^6}}             \nn && \left.
+{\frac {15{\xi}^{2}f'^{2}}{128r^2f^2}}
-{\frac {5\xi\,f'^{2}}{1024r^2f^2}}
\right]{S^0_3(\varepsilon, \mu)}_{\displaystyle|_{ \varepsilon=0}}
+\left[
\left (
{\frac {5r^2f'^{2}}{64f^2}}
-{\frac {5\xi\,r^2f'^{2}}{16f^2}}  \right ){m}^{4} 
+\left (
{\frac {5f'(r^2)'}{128r^2f}} \right. \right. \nn && \left.
-{\frac {15\xi\,f'(r^2)'}{32r^2f}}
-{\frac {5{\xi}^{2}f'^{2}}{4f^2}}
+{\frac {5{\xi}^{2}f'(r^2)'}{4r^2f}}
+{\frac {15\xi\,f'^{2}}{32f^2}}
-{\frac {5f'^{2}}{128f^2}}                      \right ){m}^{2}
+{\frac {5(r^2)'^2}{1024r^6}}                              \nn && 
+{\frac {5{\xi}^{3}f'(r^2)'}{2r^4f}} 
-{\frac {25\xi\,(r^2)'^2}{256r^6}}
-{\frac {5f'(r^2)'}{512r^4f}}
-{\frac {25\xi\,f'^{2}}{256r^2f^2}}
+{\frac {25\xi\,f'(r^2)'}{128r^4f}}
+{\frac {5{\xi}^{2}(r^2)'^2}{8r^6}}                        \nn && \left. \left.
+{\frac {5f'^{2}}{1024r^2f^2}} 
-{\frac {5{\xi}^{2}f'(r^2)'}{4r^4f}}
-{\frac {5{\xi}^{3}(r^2)'^2}{4r^6}} 
-{\frac {5{\xi}^{3}f'^{2}}{4r^2f^2}}
+{\frac {5{\xi}^{2}f'^{2}}{8r^2f^2}}
\right]{\frac{\partial^2 }{\partial \varepsilon^2}
S^0_3(\varepsilon,\mu)_{\displaystyle|_{ \varepsilon=0}}}
\right\}.
\ear



\begin{thebibliography}{99}
\bibitem{MT}
M. S. Morris and K. S. Thorne, Am. J. Phys. {\bf 56}, 395 (1988).
\bibitem{MTY}
M. S. Morris, K. S. Thorne and U. Yurtsever, \prd {\bf 61},
1446 (1988).
\bibitem{hocvis} E.E. Flanagan and R.M. Wald, \prd {\bf 54}, 6233 (1996),\\
D. Hochberg and M. Visser, \prd  {\bf 56}, 4745 (1997).
\bibitem{BW1}
C. Barcel\'o and M. Visser, Phys. Lett. {\bf B466}, 127 (1999).
\bibitem{BW2}
C. Barcel\'o and M. Visser, Class. Quantum Grav. {\bf 17}, 3843 (2000).
\bibitem{Od2}S. Nojiri, O. Obregon, S.D. Odintsov, K.E. Osetrin,
Phys.Lett. {\bf B 458}, 19 (1999).  
\bibitem{Agnese}
A. Agnese and M. La Camera, \prd {\bf 51}, 2011 (1995).
\bibitem{Nandi}
K. K. Nandi, A. Islam, and J. Evans, \prd {\bf 55}, 2497 (1997).
\bibitem{Anch}
L. A. Anchordoqui, S. Perez Bergliaffa, and D. F. Torres, \prd {\bf 55},
5226 (1997).
\bibitem{Hoch}
D. Hochberg, Phys. Lett. {\bf B251}, 349 (1990).
\bibitem{Bhawal}
B. Bhawal and S. Kar, \prd {\bf 46}, 2464 (1992).
\bibitem{Sush}
S. V. Sushkov, Phys. Lett. {\bf A164}, 33 (1992).
\bibitem{HPS}
D. Hochberg, A. Popov, and S. V. Sushkov, \prl {\bf 78}, 2050 (1997).
\bibitem{Page}
D. N. Page, \prd {\bf 25}, 1499 (1982).
\bibitem{BO}
M. R. Brown and A. C. Ottewill, \prd {\bf 31}, 
2514 (1985).
\bibitem{BOP}
M. R. Brown, A. C. Ottewill, and D. N. Page, \prd {\bf 33}, 
2840 (1986).
\bibitem{FZ}
V. P. Frolov and A. I. Zel'nikov, \prd {\bf 35}, 
3031 (1987).
\bibitem{FZ1}V. P. Frolov and A. I. Zel'nikov, Phys.~Lett.~B {\bf 115},
372 (1982).
\bibitem{FZ2}V. P. Frolov and A. I. Zel'nikov, Phys.~Lett.~B {\bf 123},
197 (1983).
\bibitem{FZ3}V. P. Frolov and A. I. Zel'nikov, \prd {\bf 29}, 1057
(1984).
\bibitem{FZ4}V. P. Frolov and A. I. Zel'nikov, \prd {\bf 35}, 3031 (1987).
\bibitem{FSZ}V. Frolov, P. Sutton and A. Zel'nikov, \prd {\bf 61}, 024021
(2000).
\bibitem{Avra1}I. G. Avramidi, hep-th/9510140.
\bibitem{Avra2}I. G. Avramidi, Teor. Mat. Fiz., {\bf 79}, 219 (1989).
\bibitem{Avra3}I. G. Avramidi, Nucl. Phys. {\bf B 355}, 712 (1991).
\bibitem{Khat1}
V. M. Khatsymovsky, Phys.~Lett. {\bf B399}, 215 (1997).
\bibitem{Khat2}
V. M. Khatsymovsky, Phys.~Lett. {\bf B403}, 203 (1997).
\bibitem{Mat}J. Matyjasek, \prd {\bf 61},124019 (2000).
\bibitem{AHS}
P. R. Anderson, W. A. Hiscock, and D. A. Samuel, \prd
{\bf 51}, 4337 (1995).
\bibitem{Sushkov2}
S. V. Sushkov, \prd {\bf 62}, 064007 (2000).
\bibitem{PS}
A. A. Popov, S. V. Sushkov, \prd
{\bf 63}, 044017 (2001).
\bibitem{CH}
P. Candelas and K. W. Howard, \prd {\bf 29}, 1618 (1984).
\bibitem{Chris}
S. M. Christensen, Phys.~Rev.~D {\bf 14}, 2490 (1976).

\end{thebibliography}
\end{document}